\documentclass[useAMS,fleqn,usenatbib]{mnras}

\usepackage{graphicx}

\usepackage[T1]{fontenc}
\usepackage{ae,aecompl}
\usepackage{hyperref}

\newcommand{\msun}{$M_\odot$}

\newcommand{\hi}{H\,{\sc i}\rm}
\newcommand{\hii}{H\,{\sc ii}\rm}

\newcommand{\nii}{[N\,{\sc ii}]}

\newcommand{\oiii}{[O\,{\sc iii}]}
\newcommand{\oii}{[O\,{\sc ii}]}

\newcommand{\oi}{[O\,{\sc i}]}

\newcommand{\sii}{[S\,{\sc ii}]}

\newcommand{\eg}{{e.g.}}
\newcommand{\ie}{{i.e.}}

\newcommand{\te}{$T_e$}

\newcommand{\lin}{$\,\lambda$}
\newcommand{\llin}{$\,\lambda\lambda$}

\newcommand{\rtf}{$r_{25}$}
\newcommand{\rd}{$r_d$}
\newcommand{\re}{$r_e$}

\newcommand{\oh}{12\,+\,log(O/H)}
\newcommand{\eo}{$\epsilon(O)$}

\newcommand{\rtwothree}{R$_{23}$}
\newcommand{\vs}{vs.}

\newcommand{\halpha}{H$\alpha$}
\newcommand{\hbeta}{H$\beta$}

\newcommand{\nodata}{...}
\newcommand{\slope}{$\nabla_{\mbox{\scriptsize O/H}}$} 
\newcommand{\slopere}{$\nabla_{\mbox{$r_e$}}$} 
\newcommand{\sloperd}{$\nabla_{\mbox{$r_d$}}$} 
\newcommand{\slopekpc}{$\nabla_{\mbox{\footnotesize kpc}}$} 
\newcommand{\slopertf}{$\nabla_{\mbox{$r_{25}$}}$} 
\newcommand{\logm}{$\log(M_\star/M_\odot)$} 

\title[Small spirals]{Metallicity gradients in small and nearby spiral galaxies}
\author[F.~Bresolin]{Fabio Bresolin\thanks{E-mail:
bresolin@ifa.hawaii.edu}\\
Institute for Astronomy, 2680 Woodlawn Drive, Honolulu, HI 96822, USA\\
}

\date{}

\pubyear{2019}

\begin{document}
\label{firstpage}
\pagerange{\pageref{firstpage}--\pageref{lastpage}}
\maketitle

%\label{firstpage}

\begin{abstract}		
\noindent
Spectra of \hii\ regions obtained with Gemini/GMOS are used to derive the radial metallicity gradients of four small, low-mass spiral galaxies. The analysis of the outer disk of one of them, NGC~1058, uncovers the characteristic flattening found in similar extended disk galaxies.
After combining these data with published long-slit observations of nearby spiral galaxies, no evidence for
a dependence of the disk scale length-normalized metallicity gradients with stellar mass is found, down to \logm\,$\sim$\,8.5.
The abundance gradients derived from these observations are 
compared to predictions from recent cosmological simulations of galaxy evolution, finding that in several cases the simulations fail to reproduce the mean steepening of the gradients, expressed in dex\,kpc$^{-1}$, with decreasing stellar mass for present-day galaxies, or do not extend to sufficiently small stellar masses for a meaningful comparison.
The mean steepening of the abundance gradients (in dex\,kpc$^{-1}$) with decreasing disk scale length is in qualitative agreement with
predictions from the inside-out model of Boissier \& Prantzos, although the predicted slopes are systematically steeper than observed. This indicates the necessity of including processes such as outflows and radial mixing in similar models of galactic chemical evolution. 
Published spatially resolved metallicity and photometric data of dwarf irregular galaxies suggest that significant, but transitory, metallicity gradients can develop for systems that have experienced recent ($t<$\,100~Myr) enhanced star formation in their inner disks.

\end{abstract}

\begin{keywords}
galaxies: abundances -- galaxies: ISM -- galaxies: spiral -- \hii\ regions.
\end{keywords}

%==============================================================================================================
\section{Introduction}

The astrophysical processes that mold the evolution of galaxies, such as stellar nucleosynthesis, gas dynamics, star formation and diffusion of metals, leave their imprint on the spatial distribution of the gas-phase metallicity 
we measure in the nearby universe.
The present-day chemical abundance distribution of an individual galaxy is the outcome of a complex pattern of evolution, advancing through merger activity, energy feedback from stars, infall of primordial gas, bursts of star formation and
metal-enriched outflows. Each of these processes, to various extents, reshuffles the metals present in and around galaxy disks.

Our knowledge of the the spatial distribution of gas phase metals (usually limited to the O/H ratio) in galaxies is predominantly expressed through the one-dimensional information provided by radial gradients.
These are commonly parameterized by the slope of the exponential metallicity drop from the center, although significant deviations occurring in the inner and outer disk regions are common (\citealt{Bresolin:2012, Sanchez-Menguiano:2018}). 
From an evolutionary standpoint this simplification is largely justified  by rapid azimuthal mixing in rotating galactic disks (\citealt{Petit:2015}), and the prevalence of those radius-dependent processes -- gas flows, star formation efficiency, inside-out disk growth -- that establish the chemical abundance gradients we observe today. 

Numerical simulations of galaxy evolution indicate that abundance gradients fluctuate at early times, due to intense starburst and merger activity, but eventually stabilize, once these disrupting events subside (\citealt{Ma:2017}). %The metallicity history of galaxies 
Remarkably, present-day isolated galaxies share a characteristic oxygen abundance gradient, if normalized to an appropriate scale length, such as the disk effective radius (\re\ -- \citealt{Diaz:1989, Sanchez:2012, Bresolin:2015}), 
the exponential scale length (\rd\ -- \citealt{Garnett:1997}), the isophotal radius (\rtf\ -- \citealt{Zaritsky:1994, Ho:2015}) or the scale length of the O/H distribution (\citealt{Sanchez-Menguiano:2018}).
\citet{Sanchez:2014} concluded that gradients normalized to \re\ share a common value, irrespective of galaxy properties, such as stellar mass, luminosity, morphology, or whether a bar is present or not.
Such a result arises naturally in galactic chemical evolution models, where the disk grows inside-out  (\citealt{Prantzos:2000, Molla:2017}). On the other hand, galaxy interactions and mergers lead to a flattening of the gradients, as a result of radial gas inflows (\citealt{Kewley:2010, Rupke:2010}).

Different conclusions regarding trends of the abundance gradients with galactic properties (\eg\ stellar mass or age) emerge from recent observational data and from simulations. For example, \citet{Belfiore:2017} and \citet{Poetrodjojo:2018} find
that low-mass galaxies have shallower gradients than massive ones, while such an effect is not detected by 
\citet{Sanchez:2014} or \citet{Sanchez-Menguiano:2016}. 
It is difficult to say whether the discrepancies are related to differences in galaxy samples, analysis and observational techniques, spatial resolution, or chemical abundance diagnostics, and possibly all these play a role.
Simulations predict different evolutionary behaviors for the slopes of the gradients. In some cases the 
gradients become shallower with time (\citealt{Pilkington:2012}), in agreement with the classical inside-out picture, 
while an increased or fluctuating strength of stellar feedback yields either flat gradients across cosmic time (\citealt{Gibson:2013}) or a considerable spread  (\citealt{Ma:2017}).

In an era increasingly dominated by Integral Field Unit (IFU) surveys of large samples of galaxies 
(\eg\ CALIFA: \citealt{Sanchez:2012}; MaNGA: \citealt{Bundy:2015}; SAMI: \citealt{Bryant:2015}), this paper follows the somewhat anachronistic approach of scrutinizing the spatially resolved chemical abundances of a moderate number of galaxies gathered from long slit data. While long slit data cannot provide competing spatial coverage of individual galaxies, current IFU studies at low redshift generally focus on significantly more distant targets, often with resolution elements that encompass regions up to several hundreds of parsecs in size. The abundance gradients thus derived can be very sensitive to corrections for seeing (\citealt{Mast:2014}) and the presence of diffuse ionized gas (\citealt{Kaplan:2016, Zhang:2017}). These issues do not affect long slit work, and abundance gradients at fairly low stellar masses, \logm\,$<$\,9, can be easily measured for nearby systems. In view of the contradicting results mentioned earlier it is worthwile to probe the conclusions drawn from long slit data of nearby objects, even though achieving restricted statistical significance.

A second motivation for this work is the comparison of the measured gradients with predictions from both simulations and chemical evolution models. Small, low-mass galaxies can play a significant role in such a comparison. For example, the 
inside-out models of \citet{Prantzos:2000} predict rather steep gradients (in physical units of dex\,kpc$^{-1}$) for spiral galaxies having scalelengths $\sim$1-2 kpc, but abundance determinations in such small galaxies are severely lacking.
This paper combines published data of nearby galaxies with new observations of four `small' (low mass and/or small scale length) late-type spiral galaxies, and compares the resulting abundance gradient properties with theoretical predictions.

Throughout this work the drop of the logarithmic oxygen abundance with radius, \slope~=~$\Delta$\,log(O/H)/$\Delta r$, \ie\ the slope of the exponential oxygen abundance gradient, will be expressed in terms of different radial units, and the abbreviations \slopere, \sloperd\ and \slopertf\
will be adopted to represent slopes in dex per disk effective radius (\re),  disk scale length (\rd), and isophotal radius (\rtf), respectively. Slopes in physical units (dex\,kpc$^{-1}$) will be indicated as \slopekpc.
Since \re\,=\,1.678\,\rd\ (\eg\ \citealt{Sanchez:2014}), we can also write \slopere\,=\,1.678\,\sloperd.

This paper is organized as follows. Sect.~2 presents new spectroscopic data of \hii\ regions in four small spiral galaxies, including nebulae located in the extended disk ($r >$\, \rtf) of NGC~1058. The oxygen abundance gradients 
of these galaxies are derived in Sect.~3. Sect.~4 introduces a sample of galaxies studied with long-slit observations, drawn from the compilation by \citet{Pilyugin:2014}. The integration of the four galaxies introduced in Sect.~2 with this larger sample constitutes the basis for comparisons with predictions made by numerical simulations and the chemical evolution models of \citet{Prantzos:2000}, carried out in Sect.~5.
The discussion in Sect.~6 focuses on abundance gradients determined in intermediate- and high-redshift galaxies, 
and concludes with considerations on the origin of abundance gradients in dwarf irregular galaxies. A summary follows in Sect.~7.

%==============================================================================================================
\section{Observations and data reduction}

In this paper data from the literature are combined with new observations of four `small' late-type spiral galaxies, characterized by exponential disk scale lengths on the order of 1-2~kpc, \ie\ placed near the bottom end of the size distribution of spiral galaxies in the nearby universe. 
One of these systems, the Sc galaxy NGC~1058, is characterized by the presence of a star-forming disk extending well beyond 
the isophotal radius. The other three are lower luminosity Sm galaxies included in the photometric studies of
\citet{Hunter:2006} and \citet{Herrmann:2013}.

The main properties of these galaxies are summarized in Table~\ref{tab:sample}, where the last column reports the 
references adopted for various parameters of the disks. 
For NGC~1058 the isophotal radius \rtf\ is taken from the RC3 catalog (\citealt{de-Vaucouleurs:1991}), while the position angle (PA) and inclination angle ($i$) are taken from \citet{Garcia-Gomez:2004}, for consistency with the compilation by \citet{Pilyugin:2014}, which will be used later in this paper.
%For the remaining galaxies these  parameters are from \citet{Hunter:2006}.
The exponential disk scale lengths (\rd) shown in Table~\ref{tab:sample}, measured in the $B$-band, are taken from \citet[NGC~1058]{Sanchez:2012} and \citet[remaining galaxies]{Herrmann:2013}. 
The distance-dependent quantities %(absolute magnitudes, stellar masses and lengths expressed in kpc) 
have been scaled by the distances assumed in this work, and summarized below. 
The stellar masses were obtained from the mass-to-light ratios calculated from the models of \citet{Bell:2001}, together with the $BV$ integrated photometry (see also Sect.~\ref{Discussion}).\\

% ..................................................................................................................
%  TABLE: SAMPLE
\begin{table*}
	\centering
	\begin{minipage}{17.5cm}
		\centering
		\caption{Gemini/GMOS galaxy sample.}\label{tab:sample}
		\begin{tabular}{cccccccccccc}
			\hline
			
			ID	& R.A.	  & Dec.	& Type	& D		& $M_B$	& log(M/M$_\odot$) & PA	& $i$	& $r_{25}$	& $r_d$	& Ref.	\\
			   	& (J2000) & (J2000)	&	    & (Mpc)	&		& 				   &	\multicolumn{2}{c}{(deg)}	& \multicolumn{2}{c}{arcsec (kpc)}	& \\
			%%(1) & (2) & (3) & (4) & (5) & (6) & (7) & (8) & (9) & (10) & (11) & (12) \\
			\hline
			NGC 1058			& 02 43 30.0	& 37 20 29  &	Sc		& 9.1	& $-$18.56 &	9.53 & 145	& 15 	& 90.6 (4.01) & 26.8 (1.18) 	& 1, a\\
			UGC 7490        	& 12 24 25.3	& 70 20 01  &	Sm		& 9.0	& $-$16.88 &	8.52 & 113	& 20		& 75.3 (3.28) & 36.8 (1.60) 	& 2, b\\
			NGC 4523        	& 12 33 48.0	& 15 10 06  &	SABm		& 16.8	& $-$17.45 &    8.64 &	24	& 35 	& 79.5 (6.49) & 30.9 (2.52) 	& 3, b\\
			NGC 4707        	& 12 48 22.9	& 51 09 53  &	Sm		& 6.5	& $-$15.73 &	8.33 & 31	& 44 	& 61.6 (1.95) & 39.8 (1.26) 	& 3, b\\[-1mm]
			\hline
		\end{tabular}
	\end{minipage}
	\begin{minipage}{17.5cm}
		$(1)$ PA, $i$ from \citet{Garcia-Gomez:2004}; \rtf\ in arcsec from \citet{de-Vaucouleurs:1991}. 
		$(2)$ PA, $i$ from \citet{Swaters:2009},  \rtf\ in arcsec from \citet{Hunter:2006}.
		$(3)$ PA, $i$, and \rtf\ in arcsec from \citet{Hunter:2006}.
		Sources for the $B$-band exponential disc scale length $r_d$ in arcsec: (a) \citet{Sanchez:2012}; (b) \citet{Herrmann:2013}. Values in brackets are expressed in kpc.\\
	\end{minipage}
\end{table*}
% ..................................................................................................................

\noindent
\textit{NGC 1058} -- the Cepheid distance to NGC~925 (9.1~Mpc; \citealt{Freedman:2001}), a galaxy in the same NGC~1023 Group as NGC~1058, is adopted.\\[-5pt]

\noindent
\textit{UGC 7490 (DDO 122)} -- I followed \citet{Kennicutt:2008}, who adopted a Local Group flow model for galaxy distances in their 11 Mpc \halpha\ survey, but using a Hubble constant $H_0 = 70~\mathrm{km\, s^{-1}\, Mpc^{-1}}$ instead of $H_0 = 75~\mathrm{km\, s^{-1}\, Mpc^{-1}}$.  This results in a distance of  9.0~Mpc.\\[-5pt]

\noindent
\textit{NGC~4523 (DDO 135)} -- 
the distance of 3.4~Mpc in \citet{Hunter:2006} and \citet{Herrmann:2013} is based on the recessional velocity (262 km\,s$^{-1}$ heliocentric), uncorrected for Virgo infall. It would imply a much better resolution into individual stars and star forming regions than actually seen in ground-based optical images of the galaxy (\eg\ from the Sloan Digital Sky Survey). 
NGC~4523 is actually considered to be a member of the Virgo Cluster (\citealt{Binggeli:1985, Kim:2014}), and the distance based on the Tully-Fisher relation, $D = 16.8$~Mpc (\citealt{Tully:1988}), is adopted. 
I note that \citet{Tikhonov:2000} assigned a much smaller distance, $D= 6.4$~Mpc, based on the photometry of the brightest stars, but with a similar method \citet{Shanks:1992} placed this galaxy at 11-15~Mpc instead.\\[-5pt]

\noindent
\textit{NGC~4707 (DDO 150)} -- the Tip of the Red Giant Branch distance of 6.5~Mpc from \citet{Jacobs:2009}, as reported in the Extragalactic Distance Database (\citealt{Tully:2009}), is adopted.\\

\begin{figure}
	\center
	\includegraphics[width=\columnwidth]{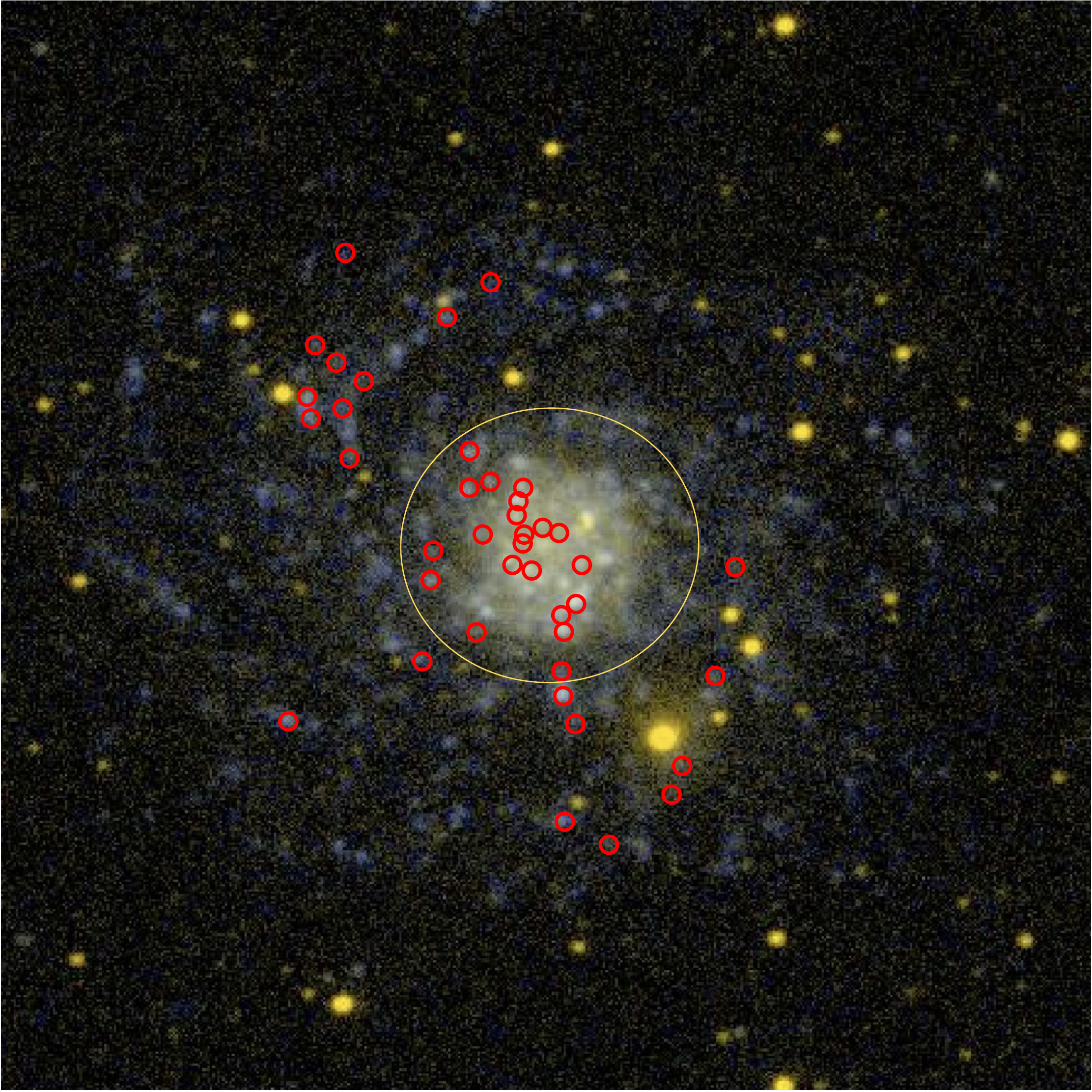}
	\caption{Far-UV {\sc Galex} color image of NGC~1058, showing the location of the observed \hii\ regions (red circles) and the projected circle of radius equal to the isophotal radius \rtf. 
		\label{galex}}
\end{figure}

Optical spectra of \hii\ regions in the four target galaxies were acquired with the
Gemini North Multi-Object Spectrograph (GMOS). The observations of NGC~1058 were designed to cover both the main optical disk of the galaxy and the extended outer disk, out to approximately 2.5 isophotal radii. Fig.~\ref{galex}
indicates the location of the \hii\ regions analysed in this work, on top of a {\sc Galex} (\citealt{Martin:2005}) far-UV color image.

% ..................................................................................................................
% ..................................................................................................................

The GMOS B600 grating was used with central wavelengths of 4500~\AA\ (`blue spectra' -- covering approximately the 3650\,-\,7100~\AA\ wavelength range) and 5800~\AA\ (`red spectra' -- approximately 4200\,-\,7100~\AA) in the case of NGC~1058, for which two partially overlapping fields were targeted. Only red spectra were secured for the other three galaxies. As a result, only for \hii\ regions in NGC~1058 was the \oii\lin3727 line observed. For all nebulae the spectral coverage includes the \oiii\lin4959+5007, \nii\lin6548+6583 and \sii\llin6717+6731 lines utilized to measure the gas metallicity (Sect.~\ref{Sec:abundances}).

The multi-object slit masks were designed based on narrow-band H$\alpha$ imaging (on- and
off-band) acquired in advance of the spectroscopic runs. The 1.4 arcsec-wide slits yielded a spectral resolution of approximately 5~\AA. Table~\ref{tab:log} summarizes dates and exposure times of the spectroscopic observations.

The raw data were reduced using {\sc iraf}\footnote{{\sc iraf} is distributed by the National Optical Astronomy Observatories, which are operated by the Association of Universities for Research in Astronomy, Inc., under cooperative agreement with the National Science Foundation.} routines contained in the {\tt\small gemini/gmos} package. The emission line intensities were measured from the flux-calibrated spectra with {\tt\small fitprofs} running under {\sc p}y{\sc raf} by fitting gaussian profiles.

The line fluxes were corrected for  interstellar extinction along the line of sight by comparing the observed \halpha/\hbeta\ ratios, corrected for a standard equivalent width of the underlying stellar population of 2~\AA, to the case B ratio of 2.86 at $10^4$~K (\citealt{Storey:1995}). The results are summarized in  Tables~\ref{tab:fluxes1058}-\ref{tab:fluxes4707} of Appendix~\ref{Appendix:B}.

% ..................................................................................................
%  TABLE: OBSERVATIONS LOG
\begin{table}
 \centering
  \caption{Observations log.}\label{tab:log}
  \begin{tabular}{lcc}
  \hline
Galaxy       & Date           & Exp. time (s) \\
\hline
NGC 1058 - Field 1		& 2011 Nov 26, 28     	&	blue: 4$\times$1800 	\\
					&						&	red: 4$\times$1200 	\\
NGC 1058 - Field 2		& 2011 Nov 29     		&	blue: 6$\times$1800 	\\
					& 2012 Jan 15, 29		&	red: 5$\times$1200 	\\
					& 2012 Feb 16, 26		&	 							\\[1mm]
UGC 7490 (DDO 122)			& 2017 Mar 29			&	red: 3$\times$1080 	\\
NGC 4523 (DDO 135)			& 2017 Jan 29			& 	red: 3$\times$1080 	\\
NGC 4707 (DDO 150)			& 2017 Jan 29, 31     	& 	red: 3$\times$1080 	\\[-1mm]
\hline
\end{tabular}
\end{table}
% ..................................................................................................

%==============================================================================================================
\section{Derivation of the oxygen abundance gradients}\label{Sec:abundances}

The spectra presented in the previous section do not uncover any of the auroral lines (such as \oiii\lin4363 and \nii\lin5755) that could provide a direct measurement of the gas temperature (\te) for the analysis of the chemical composition of the ionized gas. The oxygen abundances relative to hydrogen (`metallicities') were therefore derived using a statistical, strong-line method. 

In Sect.~\ref{Discussion} the information on the abundance gradients obtained for the four galaxies presented in Table~\ref{tab:sample} will complement similar data extracted from the compilation of nearby galaxies by \citet[=\,P14]{Pilyugin:2014}, who derived the oxygen abundances with their 
counterpart ($C$) method. This method determines the oxygen abundance, \eo\,$\equiv$\,\oh, by looking at the best 
line intensity match with a reference sample of \hii\ regions with known \te-based abundances. In cases where lines redward of \oiii\llin4959+5007 were unavailable P14 used a variant of the $P$ method (\citealt{Pilyugin:2005a}). 

Unfortunately, the $C$ method cannot be easily reproduced, since access to the full database of template spectra is required. Moreover, the \oii\lin3727 line, required to apply this technique, is missing for three of the four galaxies. In order to obtain \eo\ values  that are in principle consistent with those from the $C$ method, the $S$ method developed by \citet{Pilyugin:2016} was adopted, utilizing a combination of the \oiii\llin4959+5007, \nii\llin6548+6583 and \sii\llin6717+6731 nebular lines.
Similarly to other strong-line nebular abundance diagnostics, the $S$ method is double-valued: oxygen abundances can be distributed along either an upper or a lower branch. The selection of the appropriate branch relies on the value of the 
\nii\llin6548+6583/\halpha\ ratio, which is monotonic with O/H. The nominal value suggested by \citet{Pilyugin:2016} for the transition between upper and lower branch is log(\nii/\halpha)$_T\,=\,-0.6$, but this choice is somewhat subjective, as explained by \citet{Pilyugin:2016}. I have adopted a value of log(\nii/\halpha)$_T\,=\,-0.7$, which yields tighter radial abundance gradients for the galaxies analyzed, without introducing significant changes in the slopes (but see the case 
of NGC~4707 described in Sect.~\ref{sect:others}).
The last column of Tables~\ref{tab:fluxes1058}-\ref{tab:fluxes4707} reports the oxygen abundances thus derived.

%==============================================================================================================
\subsection{Gas metallicity in NGC~1058 and its extended disk}\label{Sec:n1058}
NGC~1058 is one of the three galaxies, together with NGC~628 and NGC~6946, where \citet{Ferguson:1998}  discovered a population of \hii\ regions located beyond the isophotal radii.  The spectroscopic analysis 
of four \hii\ regions in the outer disk of this galaxy (objects 7, 25, 35 and 40 in Table~\ref{tab:fluxes1058}) carried out by \citet{Ferguson:1998a} was, however, insufficient to characterize the overall chemical abundance of this extended disk. Subsequent spectroscopy with IFUs concentrated on the inner disk only (\citealt{Rosales-Ortega:2010, Mast:2014}). 

% ..................................................................................................................
\begin{figure}
	\center
	\includegraphics[width=\columnwidth]{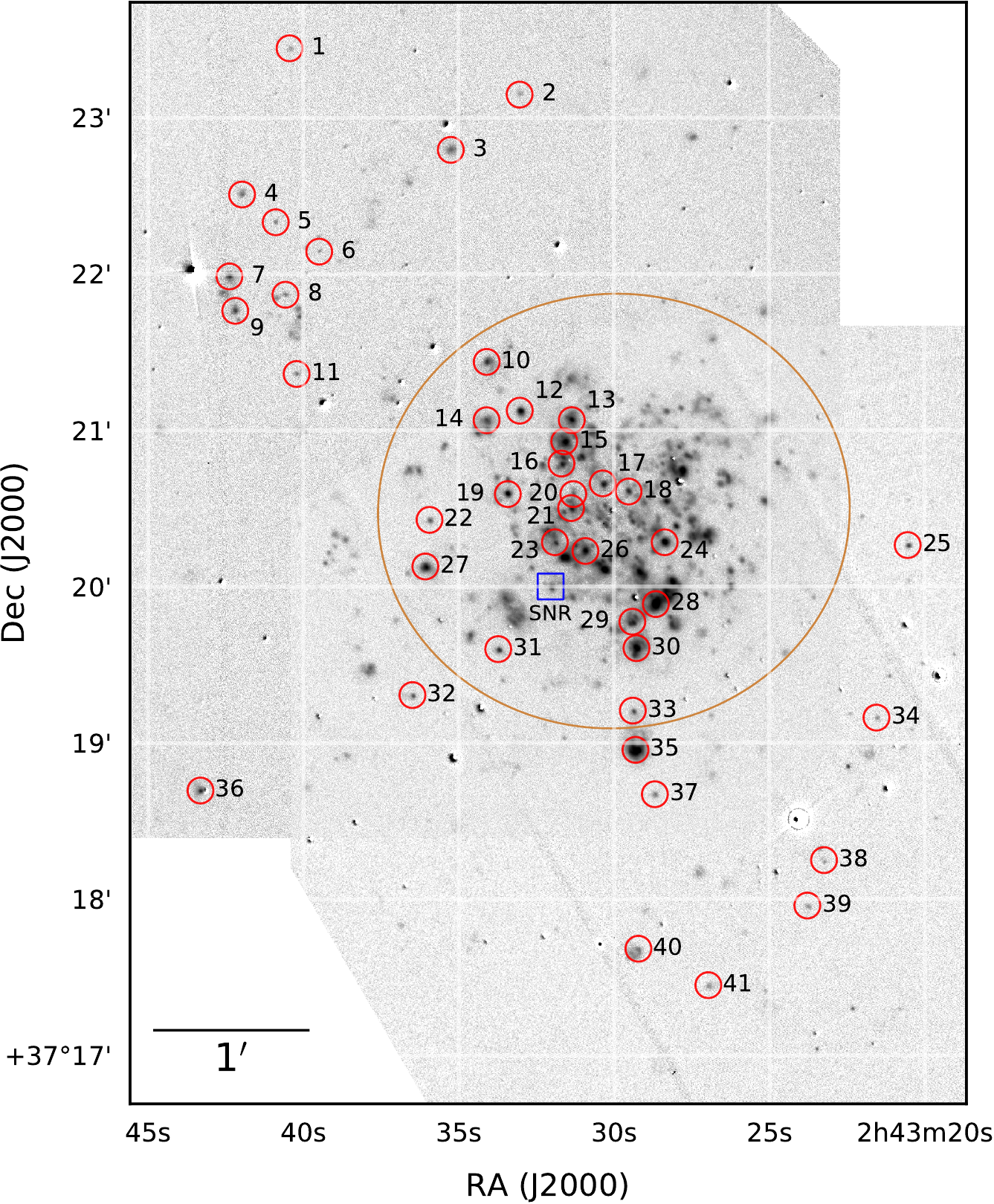}
	\caption{Identification of the \hii\ regions observed in NGC~1058. The image is a mosaic of the continuum-subtracted \halpha\ images used to prepare the GMOS slit masks. The projected circle of radius equal to the isophotal radius \rtf\ is shown. The location of a supernova remnant (SNR) candidate is marked by the blue square.
		\label{halpha}}
\end{figure}
% ..................................................................................................................

As shown in the {\sc Galex} image of Fig.~\ref{galex}, the star-forming disk of NGC~1058 is populated by numerous young stellar clusters, and extends to at least 2.5 isophotal radii (10~kpc) from the galaxy center. The spiral structure delineated by the UV-emitting young star forming regions is easily discerned. Half of the \hii\ regions observed for this work are located beyond \rtf, and are found in association with the outer spiral arms.
Fig.~\ref{halpha} identifies the \hii\ regions included in 
Table~\ref{tab:fluxes1058} on a continuum-subracted \halpha\ mosaic created from the GMOS preparatory images.
The position of a supernova remnant (SNR) candidate, identified by strong \oi\lin6300, \nii\llin6548+6583 and \sii\llin6717+6731 line emission, is marked with a square.

\begin{figure}
	\center
	\includegraphics[width=\columnwidth]{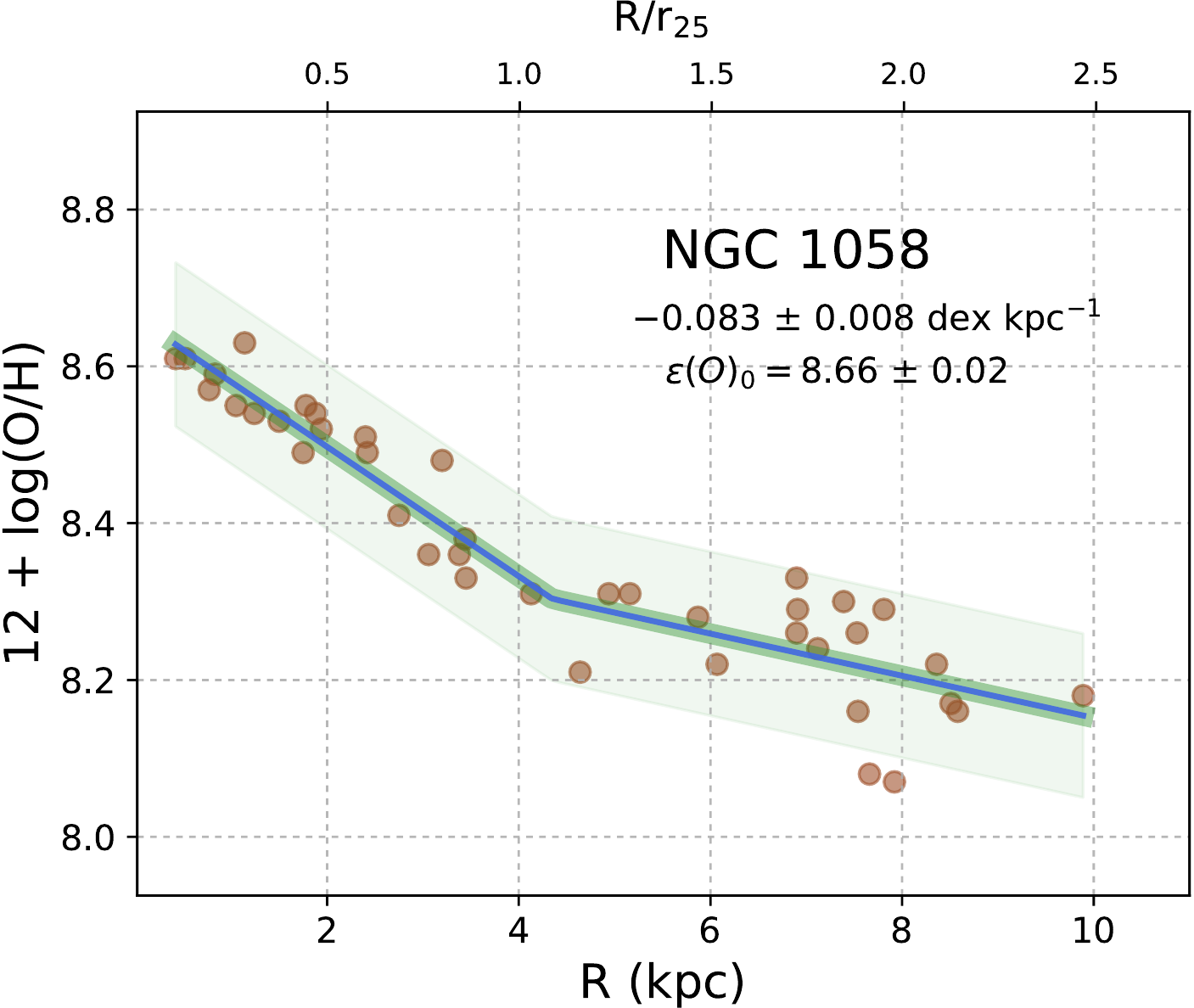}
	\caption{Oxygen abundance gradient and piecewise linear fit for NGC~1058. The shaded area represents the 95\% confidence interval. The gradient slope is quoted for the inner disk only, together with the intercept of the linear fit \eo$_0$.
		\label{n1058_gradient}}
\end{figure}

The radial O/H abundance gradient derived from the data in Table~\ref{tab:fluxes1058}, illustrated in Fig.~\ref{n1058_gradient}, features a flattening in the outer disk ($r >$ \,\rtf). 
There is no evidence for the inner disk ($r < 0.2$\,\rtf) flattening or inversion detected by the integral field spectroscopic data of \citet{Mast:2014}, possibly because of the limited statistics.

In order to determine the location of the outer metallicity break and the change of the exponential slope of the gradient following an objective procedure, as well as ensure the continuity of the function describing the radial gradient, the python library {\tt pwlf} was applied to the data. This library can be used to determine a piecewise linear least-square fit to a two-dimensional data array with a specified number of breaks, whose location can be either determined from the data or pre-defined. The best fit, shown by the line in Fig.~\ref{n1058_gradient}, yields a break at $r=4.22$~kpc (1.05\,\rtf\ = 2.13\,\re), and the slope  flattens from \slopekpc\ = $-0.083 \pm 0.008$ to \slopekpc\ = $-0.026 \pm 0.010$.
The slope of the inner abundance gradient determined by \citet{Pilyugin:2014} with the $C$ method, using line fluxes published by \citet{Ferguson:1998} and \citet{Sanchez:2012}, scaled to the distance adopted here, is $-0.088 \pm 0.007$ dex\,kpc$^{-1}$, virtually identical to what is obtained from the new Gemini data.

The behavior of the radial abundance gradient in NGC~1058 is analogous to what has been found, out to similarly large radii (2.0-2.5$\times$~\rtf), in the study of other isolated extended disk galaxies, \eg\ M83 (\citealt{Bresolin:2009}) and NGC~3621 (\citealt{Bresolin:2012}). These studies also showed mean O/H abundances of the outer disks similar to what is measured for NGC~1058, on the order of \eo\,$\simeq$~8.2\,-\,8.4.
The flattening of the outer oxygen abundance gradients is relatively common according to investigations of large samples of galaxies, such as CALIFA (\citealt{Sanchez:2014, Sanchez-Menguiano:2016}), even though it should be pointed out that such studies are mostly confined to the main optical disks. In these cases this flattening trend has not been, to my knowledge, associated with the presence of outer, extended star-forming disks. These studies find that the `breaks' occur typically at galactocentric distances equivalent to 1.5-2\,$\times$\,\re, although with a very broad distribution ($\sim$0.3-2.8\,$\times$\,\re). 
Adopting $g$-band inner disk scale lengths for CALIFA galaxies from \citet{Mendez-Abreu:2017} and \rtf\ values from the RC3 (as reported in NED\footnote{The NASA/IPAC Extragalactic Database (NED) is operated by the Jet Propulsion Laboratory, California Institute of Technology, under contract with the National Aeronautics and Space Administration.}), a mean \rtf/\re\ ratio of $1.87\, \pm\, 0.66$ is found, which makes the typical break location compatible with the idea that it occurs in the vicinities of the isophotal radius, as in the case of extended disk galaxies.

The work on the chemical abundances of extended disks carried out so far on a very small sample of spiral galaxies (summarized in \citealt{Bresolin:2017}, and including now NGC~1058)
has been able to sample the radial gradients across the optical edges of galaxies, to much larger galactocentric distances than possible for the IFU-based surveys carried out so far.
Such work indicates that if a galaxy has an outer star-forming disk then, invariably, it has a flat or nearly-flat radial abundance distribution, with a break that takes place at or around the isophotal radius \rtf, as in the case of NGC~1058 presented here. %Therefore, it is not obvious that the abundance breaks observed inside disks (\eg\ by CALIFA) and those occurring further out in the case of extended disk galaxies have the same physical cause.
Very different mechanisms can in principle be responsible for the flattening of the abundance gradients of extended disk galaxies. Broadly speaking, these can be related to a flattening of the star formation efficiency, gas mixing from radial processes and metal-enriched infall. Speculations concerning these mechanisms have been recently summarized by \citet{Bresolin:2017}, to which the reader is referred for further details.

% ..................................................................................................................
% ..................................................................................................................

%==============================================================================================================
\subsection{Gas metallicity in the low-mass spirals}\label{sect:others}

The oxygen abundances of the \hii\ regions observed in UGC~7490, NGC~4523 and NGC~4707 were derived, as described earlier,
adopting the $S$ method. In the case of NGC~4707 it was found that the choice of the transition value log(\nii/\halpha)$_T$
between upper and lower branch had a significant effect on the slope derived for the gradient. Most significantly,  for the adopted log(\nii/\halpha)$_T\,=\,-0.7$ the 
slope is quite steep, 
\slopekpc\ = $-0.27$. While slopes around $-0.20$~dex\,kpc$^{-1}$ or steeper have been reported for a handful of low-mass galaxies (\citealt{Ho:2015}), in this case this is likely a result of the inadequacy of a double-valued diagnostic such as the $S$ method to provide reliable abundances near the turnover region located between the upper  and lower branches. This is suggested by looking at the slopes one obtains adopting alternative abundance diagnostics, that are monotonic with \eo, such as 
N2 = log(\nii\lin6583/\halpha), O3N2 = log(\oiii\lin5007/\hbeta)/(\nii\lin6583/\halpha) (\citealt{Pettini:2004}) or 
the combination of \nii\lin6583 and \sii\llin6717+6731 line intensities proposed by \citet{Dopita:2016}.
For both UGC~7490 and NGC~4523 these indicators provide abundance gradients that are compatible with those obtained from the $S$ method (the \citealt{Marino:2013} calibration for N2 and O3N2 was adopted). However, for NGC~4707
the \slopekpc\ values are found to be much shallower, in the range $-0.03$ to $-0.07$, and with uncertainties comparable to the derived slope values. For the \hii\ regions in NGC~4707 I have therefore adopted the mean O/H abundances obtained from N2, O3N2 and the \citet{Dopita:2016} indicator.

\begin{figure}
	\center
	\includegraphics[width=0.95\columnwidth]{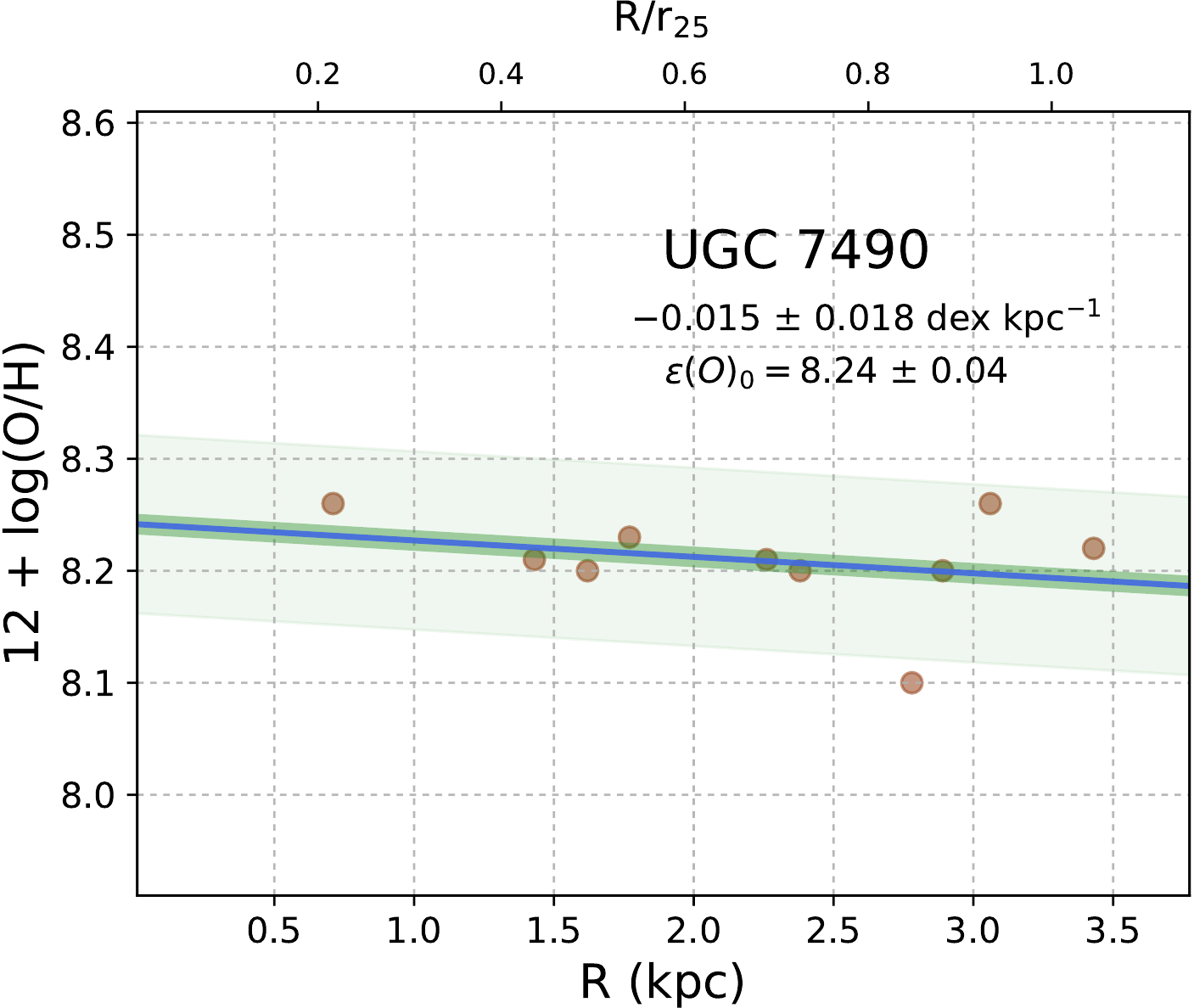}\\[5mm]
	\includegraphics[width=0.95\columnwidth]{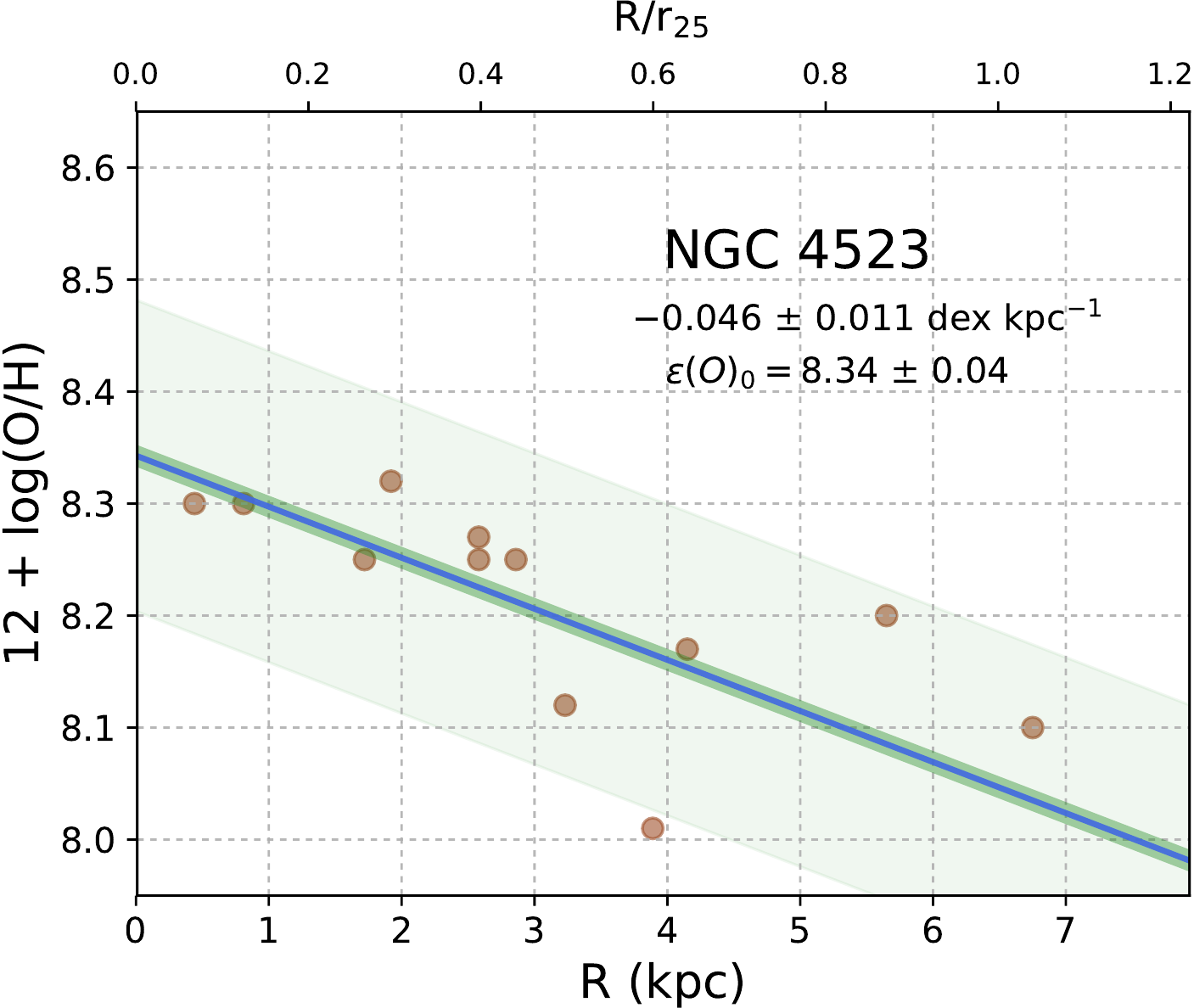}\\[5mm]
	\includegraphics[width=0.95\columnwidth]{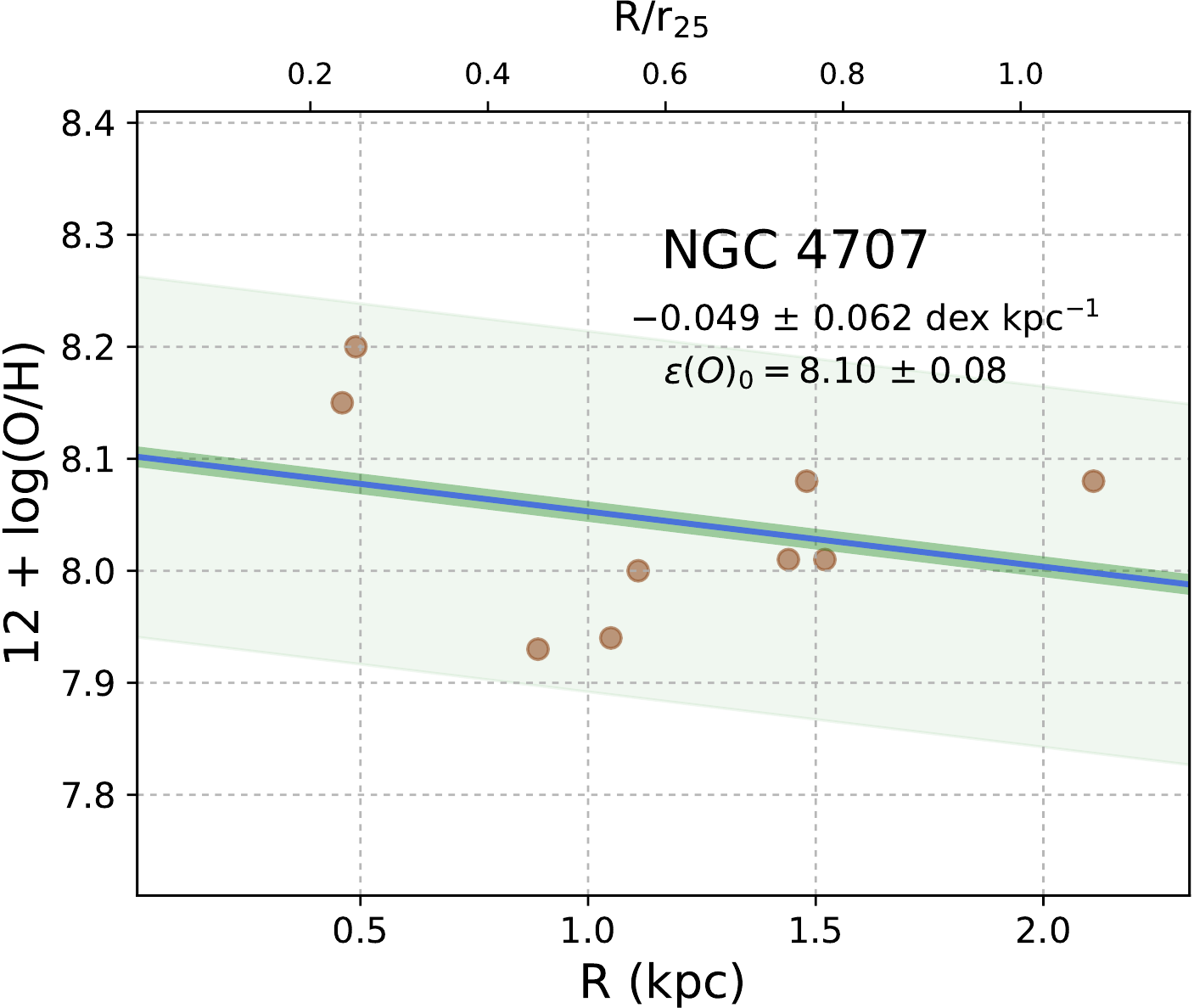}
	\caption{Oxygen abundance gradients and linear regressions for UGC~7490, NGC~4523 and NGC~4707. The shaded areas represent the 95\% confidence intervals. The values of the slope in dex\,kpc$^{-1}$ and of the intercept \eo$_0$ are shown below the galaxy name.
		\label{others_gradient}}
\end{figure}

The abundance gradients of the low-mass spirals are displayed in Fig.~\ref{others_gradient}, where both the values of the slope \slopekpc\ and the intercept \eo$_0$ are indicated. The relatively small number of \hii\ regions analyzed in these galaxies (from 9 to 13 -- however well distributed along radius) makes the gradient slopes significantly more uncertain than in the case of  NGC~1058 (where 40 nebulae were analyzed).
In two instances (UGC~7490 and NGC~4707), given the uncertainties of the linear regressions
the derived abundance gradients are consistent with being flat. The more uncertain result and largest scatter (0.08 dex rms \vs\ 0.03 dex) are found for NGC~4707, the faintest and least-massive of the galaxies considered here.

% ..................................................................................................................
% ..................................................................................................................

%==============================================================================================================
\section{Long slit sample of spiral galaxies}\label{Discussion}

This work was partly motivated by the suggestion made by \citet{Prantzos:2000} to investigate the chemical abundance gradients in galaxy disks with small scale lengths, in order to verify the predictions of a semi-analytic chemical evolution model (\citealt{Boissier:1999}).
In addition to testing model predictions with observations, with emphasis on small galaxies, it is of interest to compare the abundance gradient properties recently obtained for large galaxy samples from Integral Field Spectroscopy (IFS) and   `traditional' long slit spectroscopy. The remainder of this paper will continue to focus on the {\em slopes} of radial abundance gradients, neglecting absolute abundance values such as the central gas metallicities of galaxies. Departures from exponentials and breaks that may be occurring in the inner or outer regions of  galaxies will also be ignored.
\smallskip

A sample of nearby spiral galaxies whose gas phase metallicity has been characterized by long slit spectroscopy was assembled, by complementing the data presented in the previous sections (for small and/or low-mass galaxies -- admittedly a very small sample) with the compilation by \citet[=\,P14]{Pilyugin:2014} of 130 nearby galaxies, where the abundance gradients were determined from published data, and homogenized using their $C$ method.  

Galaxies from the P14 study were removed from further analysis if they {\em (a)} are part of interacting systems; {\em (b)} have an inclination to the line of sight $i>70$~\,degrees; 
{\em (c)} have a number of independent \hii\ region abundances $N <8$; {\em (d)} include data from IFS.
Irregular galaxies were also removed, since ultimately the comparison will be carried out for models of spiral galaxy disks. These constraints reduced the final sample to 46 galaxies (including the four galaxies analyzed earlier). NGC~1058 was included by P14 (inner disk data only), and I retained the (equivalent) inner disk gradient information extracted from the Gemini data. 

Distances  were re-evaluated based on the availability of Cepheid or Tip of the Red Giant Branch studies. For galaxies lacking these indicators  distances from the Extragalactic Distance Database (\citealt{Tully:2009}) and, further away
($D > 44$~Mpc),  NED, were used, adopting infall-corrected values and $H_0 = 70$\,km\,s$^{-1}$\,Mpc$^{-1}$. %, including the correction for the reference frame defined by Virgo+GA+Shapley)
Absorption-corrected total $B$ magnitudes and  $B-V$ colors were assembled from HyperLeda\footnote{http://leda.univ-lyon1.fr/} (\citealt{Makarov:2014}). When the corrected color was not available, a color-morphological type relation obtained from the HyperLeda database was used.
Galaxy masses were obtained from the mass-to-light ratios calculated from the models of \citet{Bell:2001} together with the $BV$ integrated photometry.
Finally, scale lengths in the $B$ or $g$  bands were extracted from the literature (in three cases the scale lengths are measured in $V$ or the near-IR).
For galaxies displaying breaks in their outer surface brightness profiles (\citealt{Pohlen:2006, Erwin:2008}) the inner disk scale length was adopted as representative of the exponential decrease of the surface brightness of the star forming disk. 
In their analysis of the surface brightness and chemical abundance properties of disk galaxies \citet{Pilyugin:2014a} find that this approximation works well, to first order.
The main properties of the sample of galaxies thus defined, henceforth referred to as the LS (Long Slit) sample, are summarized in Table~\ref{tab:galaxies}. 

A comparison IFS sample of spiral galaxies was obtained by combining the CALIFA survey data presented by \citet[=\,SM16; 122 galaxies]{Sanchez-Menguiano:2016} with the VLT/MUSE sample presented by \citet[=\,SM18; 95 galaxies]{Sanchez-Menguiano:2018}.
For consistency with the criteria used to build the LS sample, interacting galaxies were removed from SM18  (interacting or merging galaxies were already excluded by SM16). There are 212 galaxies in the IFS sample. SM16 and SM18 used the O3N2 abundance diagnostic, as calibrated by \citet{Marino:2013}, therefore the  gradient slope values they derived cannot be compared quantitatively with those computed by P14, who used the $C$ method, but this does not affect the main conclusions presented below. Additionally, it should be noted that exponential  abundance gradients are fitted across the full radial range available by P14, while SM16 and SM18 only fit abundance data that lie in the range $0.2\, <$\,r/\re\,$< 2.0$, in order to exclude potential abundance gradient breaks in the inner and outer parts of the disks.
\citet{Pilyugin:2017} found that the  difference between the two approaches is, to first order, negligible.

%==============================================================================================================
\subsection{Characterization of the Long Slit sample}\label{Sect:characterizing}

% ..................................................................................................................
\begin{figure}
	\center
	\includegraphics[width=\columnwidth]{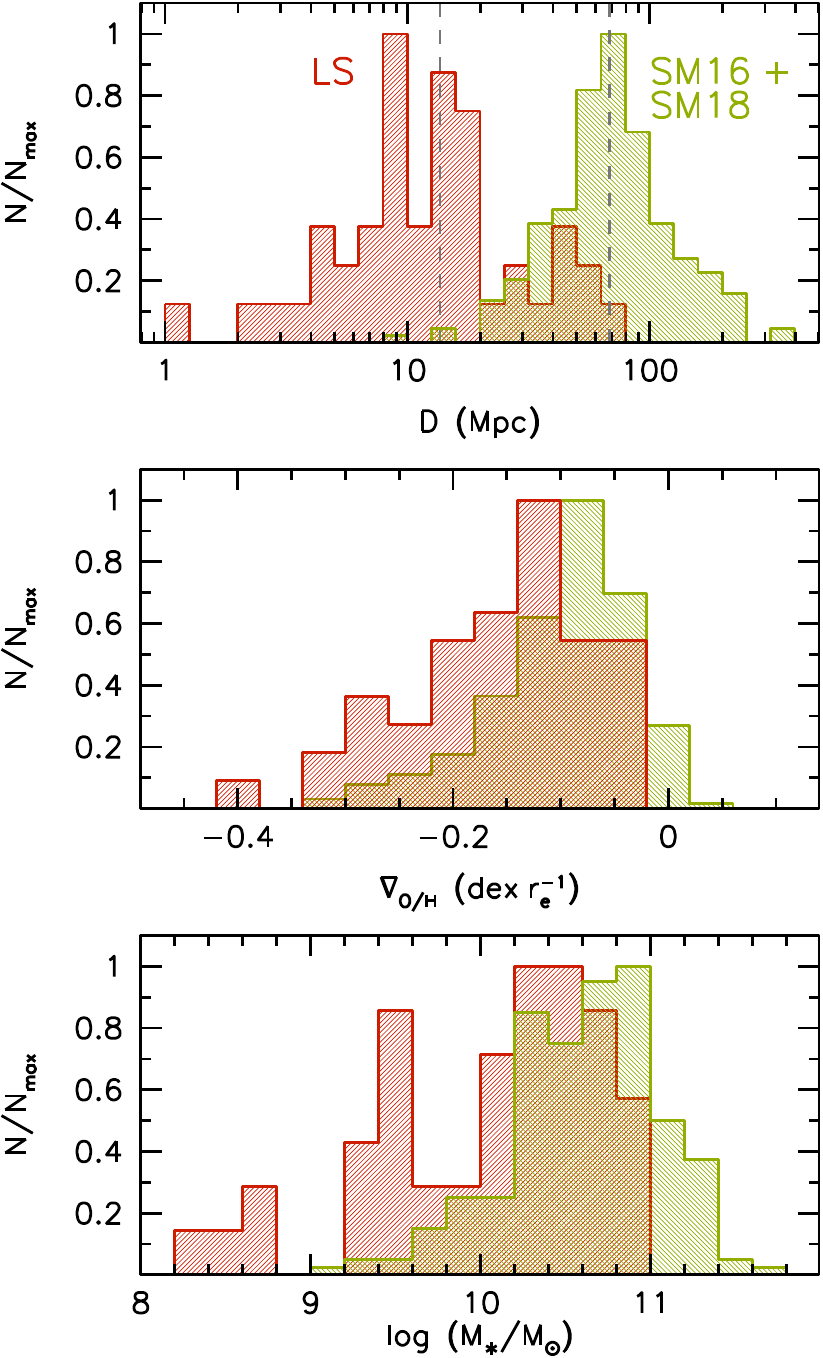}
	\caption{{\em (Top)} Normalized histograms of galaxy distances in the LS sample (in red) and in the IFS sample (in green). The median values (13.6 Mpc and 68.3 Mpc, respectively) are indicated  by the vertical lines. {\em (Middle)} Distributions of the normalized abundance gradient slopes (\slopere).
		{\em (Bottom)} Distributions of the galaxy stellar masses. \label{histogram}}
\end{figure}
% ..................................................................................................................

The LS sample represents a heterogeneous assembly of spiral galaxies observed with  single- or multi-slit spectroscopy by a variety of observers, who used different instrumentation and independent data reduction techniques. The analysis by P14 homogenized the \hii\ region abundances from the published emission line fluxes adopting a single abundance diagnostic. In this Section this sample is characterized and compared to the IFS comparison sample, with the main purpose of testing whether, despite its diverse nature, the LS sample still provides results that are in broad agreement with the comparison sample.

The galaxies in the LS sample are mostly nearby, with a median distance of 13.6~Mpc (red histogram in Fig.~\ref{histogram}, {\em top}). At this distance one arcsecond, which is the typical slit width used for slit spectroscopy, corresponds to 66~pc.
Recent IFS surveys of nearby galaxies have mainly focused on more remote targets, in order to fit the optical disks into the instrumental fields of view. Exceptions are represented by PINGS (\citealt{Rosales-Ortega:2010}) or other projects where tiling has been employed (\citealt{Sanchez:2015}), but the number of galaxies observed in this mode remains small.
The distance distribution of the IFS comparison sample, illustrated by the green histogram in  Fig.~\ref{histogram} ({\em top panel}), has a median value of 68.3~Mpc, overlapping with the large-distance tail of the
LS sample distribution in the 20-60~Mpc range.
%Considering the spatial resolution of the CALIFA dataset on the sky (3 arcsec -- \citealt{Sanchez:2014}), the median physical resolution in the IFS sample is 993~pc. YES, but SM18 uses MUSE, with better spatial resolution.

The middle panel of Fig.~\ref{histogram} displays the normalized histogram of  \slopere\  for the LS (red) and IFS (green) samples. %Despite a virtually equal range coverage in slope, the two distributions are significantly different, as confirmed by a K-S test. The LS sample contains a higher proportion of galaxies with steeper gradient slopes, and has a more symmetric distribution around the peak.
The {\em mean} gradients are \slopere\,(LS)\,=\,$-0.16 \pm 0.09$ and \slopere\,(IFS)\,=\,$-0.09 \pm 0.07$. %(the error quoted is the standard deviation). 
The value reported for non-interacting galaxies by \citet{Sanchez:2014} is $-0.11 \pm 0.08$ dex\,$r_e^{-1}$. The {\em median} gradients are \slopere\,(LS)\,=\,$-0.13 \pm 0.06$ and \slopere\,(IFS)\,=\,$-0.08 \pm 0.04$.
The small offset between the LS and IFS samples can be attributed to the use of
 different abundance diagnostics, as found in various investigations (\eg\ \citealt{Ho:2015}, SM18).

Because of its proximity the LS sample reaches lower galaxy stellar masses than the IFS sample.
This is illustrated in Fig.~\ref{histogram} ({\em bottom panel}).
Three of the four objects with \logm\, $< 9.0$ in this histogram are the result of the Gemini observations presented here. This low-mass regime has not been explored yet by IFU surveys of nearby galaxies.

%..................................................................................................................
\begin{figure*}
	\center
	\includegraphics[width=\textwidth]{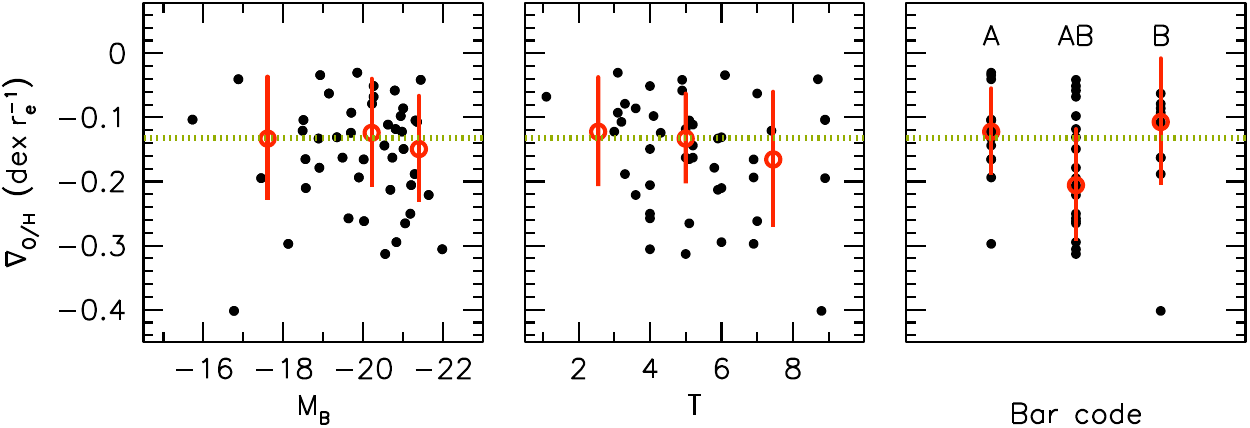}
	\caption{The O/H abundance gradient slope normalized to the disk effective radius as a function of galaxy absolute magnitude $M_B$ {\em (left)}, 
		morphological type T {\em (center)} and bar code ({\em right}; A: no bar; AB: weak bar; B: strong bar). 
		The range in both $M_B$ and T has been divided into three bins of approximately equal size.
		The open red circles represent the median slope in each bin. The horizontal dotted line indicates the median slope for the full sample 
		($-0.13 \pm 0.06$ dex\,$r_e^{-1}$, where the error is the semi-interquartile range).
		\label{threeplots}}
\end{figure*}
%..................................................................................................................

Fig.~\ref{threeplots} illustrates the slope of the abundance gradient \slopere\,  as a function of galaxy absolute magnitude ($M_B$), numerical morphological type (T) and bar code. In the case of the $M_B$ and T diagrams three bins of approximately equal size have been created, while the bar code for each galaxy is either A (no bar), AB (weak bar) or B (strong bar). The median slope value in each bin is shown by open red circles, together with its standard deviation. The median for the full sample  is represented by the dotted horizontal line. Given the size of the error bars no obvious trend is apparent when considering either $M_B$ or T, in agreement with results from SM16 and SM18,
although \citet{Perez-Montero:2016} found evidence for slightly steeper slopes in intermediate-type (Sbc) spirals compared to Sa/Sab and Sd/Irr types. 
In addition, no significant trend with the presence or absence of bars is noticed in the right panel of Fig.~\ref{threeplots}, also in agreement with recent studies (\citealt{Sanchez:2014, Kaplan:2016}, SM18).
%However, because of its size, the LS sample is not ideal to draw conclusions about the potential effect of bars on the slopes of the abundance gradients. 
Fig.~\ref{threeplots} therefore emphasizes the fact that the properties of the normalized abundance gradients of the LS sample galaxies are in general agreement with those found in the larger IFS sample chosen as a reference.

%..................................................................................................................
\begin{figure}
	\center
	\includegraphics[width=\columnwidth]{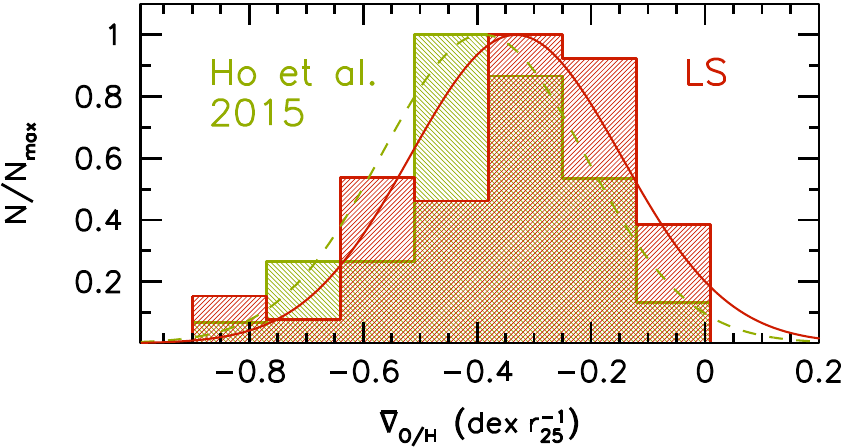}
	\caption{Normalized histograms of the normalized abundance gradients (\slopertf) of the LS sample (in red) and the
		\citet{Ho:2015} sample of 49 galaxies (in green). The gaussian curves corresponding to the mean values and standard deviations of the
		two distributions are shown by the continuous (red) and dashed (green) lines, respectively. 
		\label{Ho2015}}
\end{figure}
%..................................................................................................................

To conclude the characterization of the LS sample, Fig.~\ref{Ho2015} compares its \slopertf\ distribution (in red) with the distribution of 49 nearby galaxies studied by \citet[in green]{Ho:2015}. The mean slope from the LS sample,
\slopertf\,$=\,-0.34 \pm 0.20$, virtually matches the `benchmark' value of 
\slopertf\,$=\,-0.39 \pm 0.18$ reported by \citet{Ho:2015}. The slight offset between the distributions can, once again, be attributed to the use of different abundance diagnostics (\citealt{Ho:2015} adopted the N2O2 indicator by \citealt{Kewley:2002}).

%==============================================================================================================
\subsection{Abundance gradients \vs\ galaxy stellar masses}\label{subsec:simulations}

% ..................................................................................................................
\begin{figure*}
	\center
	\includegraphics[width=\textwidth]{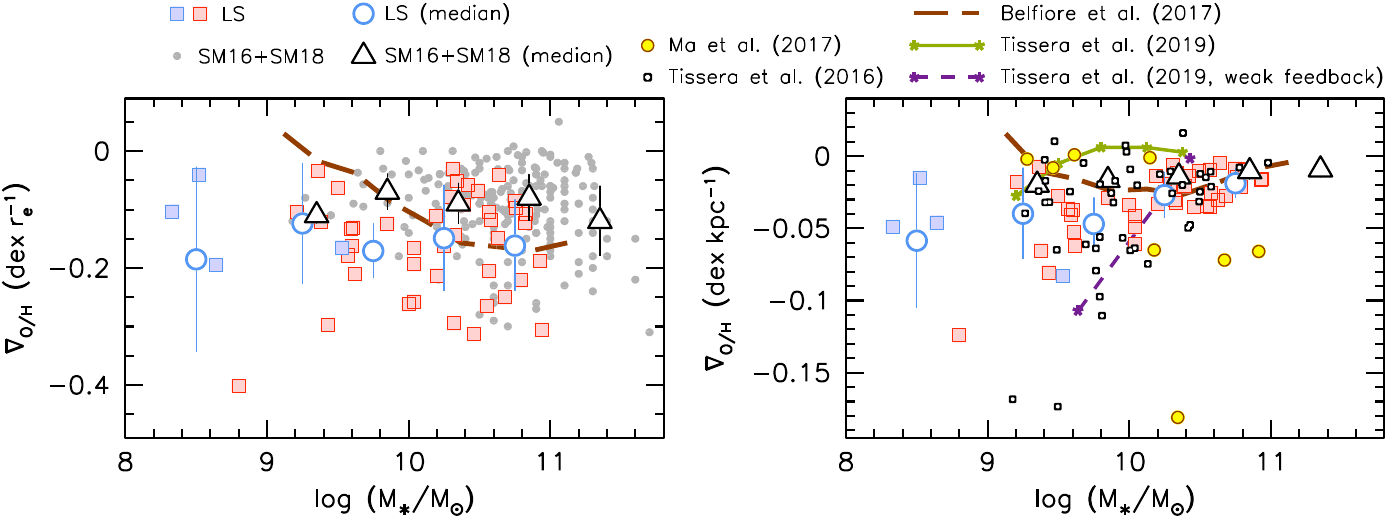}
	\caption{{\em (Left)} Normalized abundance gradient slope (\slopere) as a function of galaxy stellar mass for the LS sample (squares -- in blue for the four galaxies newly observed with Gemini) and the IFS reference sample (grey dots).  The median calculated in 0.25 dex mass bins from \citet[from their Fig.~9]{Belfiore:2017} is shown by the dashed line. {\em (Right)} Slope in dex\,kpc$^{-1}$ as a function of galaxy stellar mass for the LS sample, compared to results from recent numerical simulations, identified in the legend above the figure.
		In both panels the open circles and open triangles represent the median measurements for the LS and IFS samples in 0.5 dex mass bins, respectively.
		\label{M-vs-re}}
\end{figure*}
% ..................................................................................................................

The dependence of abundance gradient slopes on galaxy stellar mass is a matter of recent controversy, in which the different abundance diagnostics employed are likely to play a role. \citet{Sanchez:2014}, SM16 and SM18 found no effect (see also \citealt{Ho:2015}, who used the \rtf\ normalization in place of \re), while work based on the MaNGA  (\citealt{Belfiore:2017}) and the SAMI  (\citealt{Poetrodjojo:2018}) surveys reach the  conclusion that gradients steepen with increasing galaxy stellar mass, up to \logm\,$\sim$\,10.5. \citet{Lian:2018}  analyzed a sample of 773 galaxies from the MaNGA survey, but, unlike \citet{Belfiore:2017}, they do not find a significant 
change of \slope\ with stellar mass in the range $10^{9} - 10^{11}$ \msun, using two different abundance diagnostics (N2 and R23), although they do not provide values for the slopes (they find, however, that the stellar metallicity gradient has a  steeper slope for massive galaxies). Similarly, \citet{Pilyugin:2019} analyzed 147 galaxies from MaNGA, and 
their analysis does not contain evidence for strong variations of the normalized slopes with stellar mass.

In the left panel of Fig.~\ref{M-vs-re}, which refers to \slopere\ \vs\ mass,
 the mass range of the LS sample has been divided into five bins, as shown
(the lowest-mass bin contains only four galaxies). No obvious trend of the median normalized slope in each bin (open circles) with stellar mass is seen, down to \logm\,$\sim$\,8.5, although it should be pointed out that the statistics (46 objects) is not comparable to that of \citet[550 objects; \citealt{Poetrodjojo:2018}, however, who also found a mass trend, used 25 galaxies]{Belfiore:2017}. 
There is thus no indication from the nearby galaxies of the LS sample that the slope of the abundance gradient flattens with decreasing galaxy mass. The analysis of the IFS sample (grey dots in Fig.~\ref{M-vs-re}; binned median values shown by open triangles) reaches the same conclusion, as already pointed out by SM16 and SM18. \citet{Belfiore:2017} attribute the discrepancy with respect to SM16, who used CALIFA data, to the fact that observations of nearby galaxies do not sufficiently sample the population of low-mass galaxies.
This explanation is not fully satisfactory, since \citet{Perez-Montero:2016} found a small dependence of \slopere\ on galaxy stellar mass also using CALIFA galaxies, but with a different abundance diagnostic. It seems likely that both sample statistics and chemical abundance analysis techniques play a role in the determination of trends between gradient slopes and galaxy masses.

The right-hand panel of Fig.~\ref{M-vs-re} expresses the slope of the abundance gradients in units of dex\,kpc$^{-1}$, and shows again that the analysis of MaNGA data by \citet{Belfiore:2017} yields flatter slopes with decreasing stellar mass below \logm\,$\sim$\,10.5 (the same authors warn about potential beam-smearing effects using non-normalized slopes). The LS data suggest the opposite trend, \ie\ a mean steepening of the gradients towards smaller masses. 
In order to quantify the observational result, I followed \citet{Ho:2015}, and split the LS sample at \logm\,=\,10.25 into two bins of equal size, and
calculated the bootstrapped mean and standard deviation for the slopes expressed in dex\,kpc$^{-1}$ for the two bins. 
For the low-mass bin I obtained a mean slope \slopekpc\,=\,$-0.043 \pm 0.005$ with a standard deviation of $0.025 \pm 0.005$ dex\,kpc$^{-1}$, while 
for the high-mass bin the mean slope is 	
\slopekpc\, =\, $-0.019 \pm 0.002$ with a standard deviation of $0.010 \pm 0.001$ dex\,kpc$^{-1}$.	
The mean slopes differ by 4.2\,$\sigma$, and the standard deviations by 3.0\,$\sigma$: the low mass galaxies in the LS sample have significantly steeper gradient slopes, as well as a higher scatter, than the massive ones, the same conclusion
reached by \citet{Ho:2015}.
It is worth noting here that a study by \cite{Hidalgo-Gamez:2011} reports \slopekpc\ $\sim$ $-0.2$ to $-0.4$ (with considerable uncertaities) for four dwarf spirals in the mass range \logm\,=\,7.0\,--\,8.1, if masses from \citet{Chang:2015} are adopted.

There is an increasing scatter towards lower masses in the right-hand panel of Fig.~\ref{M-vs-re}. Such an effect is well-known from  previous studies (\eg\  \citealt{Garnett:1997, Ho:2015}). \citet{Carton:2018} obtained a similar result for intermediate-redshift ($0.08 < z < 0.84$) galaxies.
The interpretation of this trend  will be provided  in Sect.~\ref{Sec:PB00}.

%==============================================================================================================
\section{Comparison with theoretical predictions}\label{modelcomparison}

\subsection{Numerical simulations}\label{subsec:simulations}

The right-hand panel of Fig.~\ref{M-vs-re} includes predictions from recent hydrodynamic simulations from \citet[yellow dots]{Ma:2017}, \citet[grey squares]{Tissera:2016} and \citet[green and magenta lines]{Tissera:2019}, calculated for present-day galaxies and covering the mass range 9.2 $<$\,\logm\,$<$ 11. 

The results from \citet[green continuous line in Fig.~\ref{M-vs-re}]{Tissera:2019} were obtained as part of the EAGLE project (\citealt{Schaye:2015}) at redshift $z=0$, and refer to their high-resolution simulation (25 Mpc box size). %, which extends to lower masses ($\log M/$\msun = 9.2) compared to the other simulation they presented (100 Mpc box size). 
Qualitatively they reproduce the increasing slope with decreasing mass that is present in the data, but the predicted slopes are exceedingly flat. Increasing the energy feedback in hydrodynamic simulations 
shifts metals from the galactic central regions into the halos (\citealt{Vogelsberger:2013}),  flattening the abundance gradients (\citealt{Gibson:2013}). \citet{Tissera:2019} make use of the EAGLE set of simulations, which provides runs with the same initial conditions but different feedback energies.
Their simulated galaxy with weak feedback at \logm\,$\sim$\,9.6 and \slopekpc~=~$-0.11$ (end point of the magenta dashed line in Fig.~\ref{M-vs-re}) indicates in fact that a better match with the data can be obtained by lowering the
feedback at small stellar masses.

The FIRE simulations (\citealt{Hopkins:2014}) resolve the feedback on sub-kpc scales. The work based on these simulations by \citet{Ma:2017} presents gradients that can display steep slopes at $z=0$, but only for \logm\,$>$\,10 (yellow circles in the right-hand panel of Fig.~\ref{M-vs-re}). The increased feedback predicted for the lower mass galaxies  mixes the metals, producing flat gradients as a result, similarly to the normal feedback case by \citet{Tissera:2019}. This is again contrary to the observations presented here (however there is agreement with the data presented by \citealt{Belfiore:2017}).

Numerical simulations calculated by \citet{Tissera:2016} adopting different sub-grid physics prescriptions are shown by grey open squares in Fig.~\ref{M-vs-re}. These results appear in better agreement with the LS data, reproducing the increased scatter for smaller galaxies. Other numerical simulations not shown here (\eg\ \citealt{Few:2012, Aumer:2013}) also indicate a steepening trend of \slopekpc\ with decreasing halo or stellar mass, as observed here.

This brief comparison between observed abundance gradients and predictions for present-day galaxies made by cosmological simulations highlights the presence of important discrepancies in some cases, especially for small systems, as well as the need to extend the simulations to lower stellar masses (\logm\ $<$ 9) at $z=0$.

%==============================================================================================================
\subsection{Boissier \& Prantzos models}\label{Sec:PB00}

In this section I  compare the LS observational data with the chemical evolution model predictions made by \citet[=PB00]{Prantzos:2000}. The models of \citet{Boissier:1999, Boissier:2000}, upon which the results presented by PB00 are based, have been adopted in several recent interpretations of spectrophotometric measurements of galaxies 
(\eg\ \citealt{Munoz-Mateos:2011, Boselli:2014, Boselli:2016, Bouquin:2015, Bouquin:2018, Fossati:2018}). PB00 presented 
model galaxy properties that can be easily related to observed quantities of star-forming disk galaxies, such as rotational velocities, sizes, colors, magnitudes and oxygen abundances.
Predictions concerning galaxy abundance gradients have been compared to observations by PB00, and more recently by  \citet{Munoz-Mateos:2011}, \cite{Yuan:2011}, \citet{Bresolin:2015} and \citet{Kaplan:2016}.

In the multi-zone model of \citet{Boissier:1999}, similarly to other chemical evolution models (\eg\ \citealt{Chiappini:2001, Molla:2005}), the spectrophotometric and chemical abundance gradients that are observed in disk galaxies stem from the prescribed radial dependence of the star formation rate ({\em SFR}) and of the gas infall rate timescale, that increases with radius and dictates the inside-out growth of galaxy disks (\citealt{Matteucci:1989}). 
Radial gas flows (\eg\ \citealt{Lacey:1985}) or gas outflows are ignored in the Boissier \& Prantzos models.
The predictions are in good agreement with a number of observables in the Milky Way. 
The model was generalized to other spiral galaxies by \citet{Boissier:2000}, who adopted simple scaling relations derived from the cold dark matter models of \citet{Mo:1998}, using only two free parameters: the spin parameter $\lambda$, proportional to the angular momentum of the galactic halos, and the maximum  velocity of the rotation curve $V_C$. The dimensionless spin parameter essentially determines the timescale of the star formation activity and therefore the extent (scale length) of the disks. 
In the models presented by \citet{Boissier:2000} $\lambda$ ranges from 0.01 to 0.09, with $\lambda = 0.03$ appropriate for the Milky Way. 
The second parameter $V_C$ affects the star formation rate and the disk mass, ranging between 80 and 360 km\,s$^{-1}$. Several observables of present-day galaxies, including the multi-wavelength Tully-Fisher relation, scale lengths,  global colors and metallicities, are reasonably well reproduced from the simple scaling adopted and the range in $\lambda$ and $V_C$  considered.

Surface brightness profiles of galaxies at multiple wavelengths (far UV to near IR)  have been fitted to a finer grid of the \citet{Boissier:2000} models by \citet{Munoz-Mateos:2011} and \cite{Bouquin:2018}. 
These authors found that the theoretical $V_C$ values derived from the fits are in relatively good agreement with the rotation velocities of spiral galaxies estimated from the \hi\ 21\,cm line widths, albeit with a tendency of over-predicting the observational data. % (errors in the inclination correction for the rotational data likely play a role in explaining the difference).

The radial chemical abundance profiles of model disk galaxies emerging from the Boissier and Prantzos framework have been discussed by PB00. Their work provided a theoretical foundation to the notion of homologous chemical evolution of galaxies emerging from the analysis of abundance gradient data by \citet{Garnett:1997}, \ie\ the similarity among galactic abundance gradients when these are normalized to the disk scale length $r_d$, or equivalently the exponential effective radius \re\ (as shown here in Fig.~\ref{M-vs-re}, {\em left}). 
The steeper abundance gradient slopes expressed in physical units (dex\,kpc$^{-1}$) of present-day lower-mass (lower-luminosity) galaxies in Fig.~\ref{M-vs-re} ({\em right}) is an effect of galaxy size, resulting from downsizing of the inside-out disk growth of galaxy disks (massive galaxies evolve first).
Variations in the spin parameter $\lambda$ are expected to
introduce a significant scatter in the abundance gradient slope of the fainter, lower mass galaxies, as observed (Fig.~\ref{M-vs-re}, {\em right}), although \citet{Bresolin:2015} pointed out that 
the expectation of flatter slopes for low surface brightness galaxies, owing to their large $\lambda$ values, is not fully confirmed by chemical abundance measurements. As seen in Sect.~\ref{subsec:simulations}, in hydrodynamic simulations the scatter arises from feedback-related mixing mechanisms.

% ..................................................................................................................
\begin{figure*}
	\center
	\includegraphics[width=0.95\textwidth]{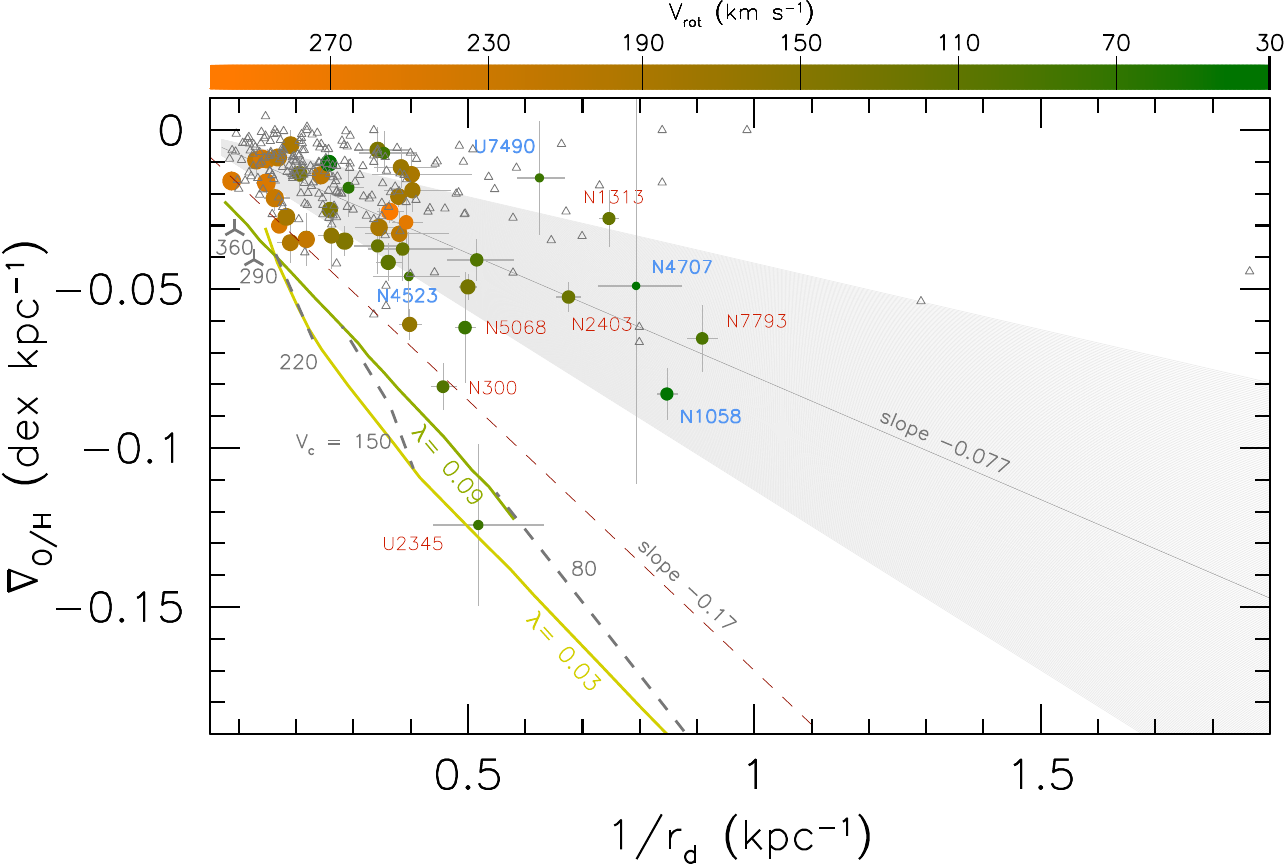}
	\caption{\slope\ (in dex\,kpc$^{-1}$) of the LS galaxies (dots) as a function of 1/\rd\ (\rd\ is the $B$-band exponential scale length). The color coding provides the rotational velocities derived from the \hi\ 21\,cm line widths. The comparison IFS sample is shown by the open triangles (scale lengths in the $g$ band). The \citet{Prantzos:2000} model results calculated for two spin parameter values ($\lambda = 0.03, 0.09$) and the full range of circular velocities $V_C$ (80-360~km\,s$^{-1}$) are shown for comparison. The slope of the dashed line ($-0.17$) corresponds to the toy model by \citet{Ma:2017}.
		The median slope in normalized units (\slopere) found for the LS sample corresponds to the line of slope $-0.077$ in this diagram, while the shaded area corresponds to the 1$\sigma$ range of the median.
		Some of the smallest galaxies are identified, including the four galaxies newly observed for this work (blue labels). The size of the LS data points is proportional to the galaxy mass.
		\label{rd}}
\end{figure*}
% ..................................................................................................................

A prediction made by PB00, and a reflection of the homologous behavior of galactic abundance gradients, 
is the existence of a tight correlation between abundance gradient slopes expressed in dex\,kpc$^{-1}$ and the inverse of the $B$-band scale length 1/\rd, with little dependence on either $\lambda$ or $V_C$. 
While the limited amount of observational data available to PB00 appeared to be in relatively good agreement with this expectation, these authors suggested to focus on the chemical abundance properties of small disk galaxies (large 1/\rd\ ratio) in order to test the models.

The models by PB00 are compared with the data from the LS sample (data points with error bars), and the comparison IFS sample (grey triangles) in Fig.~\ref{rd}. The figure includes model predictions (green curves) for 
the two extremes of the range in spin parameter, $\lambda=0.03$ and $\lambda=0.09$ (since $\lambda = 0.01$ was considered non-physical), showing the relatively small effect of the spin parameter in this diagram, and considers the full range of the model $V_C$ values (80, 150, 220, 290 and 360~km\,s$^{-1}$). Rotational velocities were gathered from HyperLeda (estimated from the \hi\ line widths), but corrected adopting the same inclination angle used to deproject the radial abundance gradients (from \citealt{Pilyugin:2014}). The color code in Fig.~\ref{rd} indicates the rotational velocities thus derived, and suggests a rough agreement with PB00, in the sense that as the rotational velocity decreases both 1/\rd\ and the absolute value of \slopekpc\ tend to increase.

Fig.~\ref{rd} includes the variation of \slopekpc\ ($-0.17$/\rd\ 
dex\,kpc$^{-1}$), according to the toy model of \citet[dashed line -- the zeropoint is arbitrary]{Ma:2017}. These authors assumed a radial exponential distribution of pristine gas and a star formation rate given by the Kennicutt-Schmidt law, together with an absence of radial mixing. It can be seen that the variation of \slopekpc\ with 1/\rd\ given by this extremely simple model is roughly consistent with that of the PB00 model curves.

In the diagram of Fig.~\ref{rd} a constant value of the slope \sloperd\ (or equivalently \slopere\ -- see Fig.~\ref{threeplots}) translates into a straight line of slope equal to the numeric value of \sloperd\ (=\,\slopere/1.678).
As seen in Sect.~\ref{Sect:characterizing}, the median slope for the LS sample is \slopere~=~$-0.13 \pm 0.06$, which corresponds to the line of slope $-0.077$ drawn in Fig.~\ref{rd}. The shaded area corresponds to
the 1\,$\sigma$ range around the median. 
Such a scatter for the normalized gradients has been attributed to differences in star formation efficiency among galaxies (\citealt{Molla:2017}) or the effects of merger history ({\citealt{Few:2012}).

It is worth pointing out that a constant scatter in \slopere\ (or \sloperd) translates into a higher scatter in \slopekpc\ for small galaxies compared to large galaxies (corresponding to the widening of the shaded area with increasing 1/\rd). This  explains why \citet{Carton:2018} found that the abundance gradient slopes of small galaxies ($r < 3$\,kpc) have a significantly higher scatter than larger galaxies when measured in units of  dex\,kpc$^{-1}$, but that the two subsamples have similar scatter  using  units of dex\,\rd$^{-1}$ (their Table~4).

Fig.~\ref{rd} indicates that the measured slopes of the LS sample galaxies are flatter than the models (except for the low-mass galaxy UGC~2345), even more so for the IFS sample (the adoption of a different diagnostic, O3N2, produced systematically flatter gradients compared to the LS sample -- as seen in Sect.~\ref{Sect:characterizing}). This could simply  result from a systematic issue arising from the abundance diagnostic adopted for the LS sample, but the same result was obtained by \citet{Munoz-Mateos:2011}, who used abundances of 21 galaxies  determined by \citet{Moustakas:2010} with the \citet{Kobulnicky:2004} theoretical calibration of \rtwothree, and by \citet{Kaplan:2016} for eight galaxies analyzed with the N2 diagnostic of \citet{Pettini:2004}. \citet{Sanchez-Menguiano:2018} showed that using the \citet{Dopita:2016} calibration yields a mean \slopere = $-0.23$ for their sample, corresponding to a line of slope $-0.14$ in Fig.~\ref{rd}, still flatter than indicated by the models. It appears that strong-line abundance indicators
currently adopted by various authors, at least those mentioned above,  do not  provide chemical abundance gradient slopes that agree with the PB00 models.

The offset between  data  and  models in Fig.~\ref{rd} could also be at first interpreted by galaxy sizes that are systematically overestimated by the models. However, if the normalized gradients are invariant, for a reduction in galaxy size there would be a corresponding steepening of \slopekpc, and the offset would remain.

The tentative conclusion is that the observed \slopekpc\ values are shallower than predicted by the PB00 models, unless the strong-line diagnostics currently available consistently underestimate the slope of the abundance gradients.
While this seems possible for diagnostics that ignore the effects of the ionization parameter, such as N2 and O3N2 
(\citealt{Pilyugin:2019}), some of the remaining strong-line methods  should be relatively free of such issues.

A systematic offset might also point to uncertainties in the stellar yields adopted in the Boissier and Prantzos models (\citealt{Munoz-Mateos:2011}) or the absence of radial gas flows in the models ({\citealt{Kaplan:2016}).
Furthermore, the effects of stellar feedback (Sect.~\ref{subsec:simulations}) and outflows are neglected in these simple models.
Chemical evolution models based on the inside-out paradigm, but including gas radial flows (\citealt{Spitoni:2011})
or outflows (\citealt{Belfiore:2019}) underscore the importance of these effects on galaxy abundance gradients.
For example, \citet{Belfiore:2019} explain how flatter gradients can be attained by increasing the outflow loading factor or lowering the infall timescale in the outer disk.

%==============================================================================================================
\section{Discussion}\label{Summary}
The downsizing assumption made in chemical evolution models (\eg\ \citealt{Boissier:2000, Molla:2005}) leads to a galaxy mass dependence of the star formation timescale: low-mass galaxies evolve more slowly and produce their stars later compared to massive galaxies. The inside-out disk growth leads then low-mass  systems to have, on average, steeper present-day abundance gradients than massive ones. The latter  developed flatter gradients thanks to the faster disk assembly. The trend observed in Fig.~\ref{rd} captures this expected behavior. According to the chemical evolution models presented by \citet{Molla:2017}, small spiral galaxies (\re \,$\sim$ 2~kpc)
started evolving their disks at $z<2$, at a time when the gradients of massive galaxies were already virtually flat. The same models also predict that the normalized slope \slopere\ does not change significantly as the disks grow, and that its value for present-day galaxies does not depend on stellar mass, as observed (\citealt{Sanchez:2014}), provided that the efficiency of star formation is `high'.

A major goal of observations, simulations and chemical evolution models of galaxies is in fact to understand how the 
chemical abundance gradients evolve with cosmic time. In recent years several investigations of the metal content of intermediate- and high-redshift galaxies, up to $z \sim 3.4$ have been carried out with seeing-limited and adaptive optics-assisted IFU observations (\eg\ \citealt{Queyrel:2012, Troncoso:2014, Forster-Schreiber:2018}), or using gravitationally lensed systems (\eg\ \citealt{Yuan:2011, Jones:2013, Wang:2017}), where spatial resolutions of only a few hundred pc can be achieved. 

The data gathered so far yield a wide variety of gradients, ranging from steep ($\simeq-0.3$ dex\,kpc$^{-1}$ -- but these extreme values are rare) to flat or positive (`inverted'), with considerable scatter at any redshift (\citealt{Carton:2018}). 
As an example, the average slope derived for $z \sim 2$ galaxies studied with adaptive optics 
by \citet{Forster-Schreiber:2018} is
$-0.030$ dex\,kpc$^{-1}$ with a scatter of $0.066$ dex\,kpc$^{-1}$
(if using the \citealt{Pettini:2004} N2 calibration).
A diversity of gradient slopes is also obtained considering only observations of lensed systems (\citealt{Leethochawalit:2016}), that arguably provide the more reliable results thanks to their higher physical spatial resolution.

Gradients with flat or positive slopes at high $z$ have been interpreted as due to the effects of cold gas inflows (\citealt{Cresci:2010}), gravitational interactions (\citealt{Stott:2014}) or feedback-generated outflows and mixing  (\citealt{Wuyts:2016}), although poor resolution imposed by seeing-limited data
or the prevalent use of the N2 indicator, which is sensitive to the N/O abundance and the ionization parameter, can introduce systematics (\citealt{Yuan:2013a, Wuyts:2016}).
The variety of slope gradients has been attributed to variations in the importance of
feedback by \citet[see also Sect.~\ref{subsec:simulations}]{Leethochawalit:2016}. \citet{Pilkington:2012}  pointed out that flatter gradients can also result from lower thresholds for star formation and high star formation efficiencies across  galaxy disks (see also \citealt{Molla:2017, Tissera:2019}), so in general the observed diversity can be expected from modifications of all these parameters.

The evolution of the abundance gradients with time predicted by chemical evolution models and cosmological simulations has been compared with observational data in several of the investigations mentioned above (see also \citealt{Swinbank:2012, Pilkington:2012, Jones:2015, Wuyts:2016, Kaplan:2016, Ma:2017, Wang:2017}). 
This has involved  predictions from a number of different cosmological simulations, including 
MUGS (\citealt{Stinson:2010}), RaDES (\citealt{Few:2012}), MaGICC (\citealt{Stinson:2013}) and FIRE (\citealt{Hopkins:2014}), as well as from suites of chemical evolution models (\citealt{Chiappini:2001, Molla:2019})

These comparisons have highlighted the importance of feedback in regulating galactic abundance gradients in simulations (\citealt{Gibson:2013}).
The adoption of an `enhanced' feedback prescription in the MaGICC simulations compared to the MUGS computations leads to virtually flat abundance gradients at all redshifts, instead of a relatively strong flattening trend between high and low $z$, as in the case of the MUGS galaxies.
This has led many authors to conclude that the generally flat gradients observed in high redshift galaxies favor the 
`enhanced' feedback solutions. The FIRE simulations shown by \citet{Ma:2017} display a variety of slopes, in good agreement with the observed scatter as a function of redshift, thanks to variable feedback during the course of the evolution of galaxies (on timescales of $10^8-10^9$ yr).

\citet{Wang:2017} have shown how the gradients correlate with stellar mass, using sub-kpc resolution data obtained from gravitationally lensed systems at redshifts $z\sim 1$ to 2.5. The sensitivity of {\em Hubble Space Telescope} grism observations allowed them to measure gradients for galaxy stellar masses down to \logm\,$\sim$\,8.0.
The intermediate-redshift sample studied by \citet{Carton:2018} extends down to an even lower mass limit, but their data is seeing-limited and does not reach sub-kpc resolution. The spatial resolution is key in removing the potential for a systematic flattening of the measured gradients (\citealt{Yuan:2013a}).

Fig.~\ref{wang} displays the \slopekpc\ \vs\ log($M$/\msun) diagram using the high-$z$ lensed data from \citet[triangles]{Wang:2017}, after removal of kinematically disturbed (non-isolated) systems, compared to the LS sample data (circles) and \citet[squares]{Ho:2015}, used as a local reference for isolated galaxies. 
Despite the larger scatter, the high-$z$ data are in substantial agreement with the local galaxies, in the sense that low-mass galaxies appear to have, on average, steeper gradients (\citealt{Wang:2017} use the word `tentative' in describing the trend -- clearly more data are required to confirm it). 

These results, combined with the comparisons presented in Sect.~\ref{subsec:simulations}, reinforce the concept that numerical simulations with revised sub-grid ingredients should be calculated, in order to reproduce
both the observed steepening of the gradients and the enhanced scatter towards low stellar masses of present-day galaxies.
The currently available simulations focus on relatively higher masses, since 
most high-redshift investigations gather data for galaxies at \logm\,$>$\,9.5 (\citealt{Forster-Schreiber:2018} and references therein). An extension to low stellar masses (down to log(M/\msun) $\sim$ 8.0) for present-day galaxies would  be important for appropriate comparisons with empirical determinations of the abundance gradients. 
It can be pointed out that observational tests of these aspects of the numerical simulations
do not necessarily require (difficult) observations of high-$z$ galaxies, but that they can also be carried out in low-mass disks in the nearby universe.

% ..................................................................................................................
\begin{figure}
	\center
	\includegraphics[width=1\columnwidth]{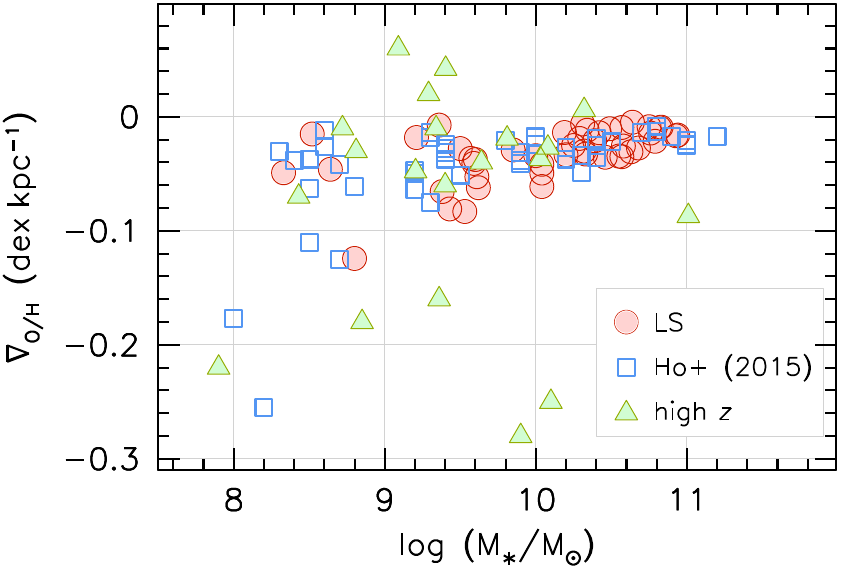}
	\caption{Abundance gradient slope as a function of \logm\ for the LS (circles) and \citet[squares]{Ho:2015} galaxies, compared to the isolated lensed systems presented by \citet[triangles]{Wang:2017}. The latter include galaxies observed by those authors, \citet{Swinbank:2012}, \citet{Jones:2013} and \citet{Leethochawalit:2016}.
		 \label{wang}}
\end{figure}
% ..................................................................................................................

\subsection{Abundance gradients in irregular galaxies}
The steepening of the mean abundance gradient with decreasing stellar mass found for spiral galaxies and the general absence of metallicity gradients in low-mass irregular galaxies (\citealt{Kobulnicky:1997a, van-Zee:2006a, Croxall:2009, Haurberg:2013}) imply some sort of transition between two different regimes in the radial distribution of metals. This was pointed out by \citet{Edmunds:1993}, who noticed the simultaneous disappearance of abundance gradients and spiral structure below an absolute magnitude $M_B \simeq -17$. This led them to conclude that abundance gradients in gas-rich galaxies are related to the presence of spiral structure.

Fig.~\ref{irregulars} revisits this concept, by showing the relationship between slope \slopekpc\ and absolute magnitude $M_B$  for the galaxies in the LS sample (red squares), \citet[blue squares]{Ho:2015} and irregular galaxies (orange dots), using data for NGC~6822 (\citealt{Lee:2006b}), NGC~1705 (\citealt{Annibali:2015}), NGC~4449 (\citealt{Annibali:2017}),
DDO~68 (\citealt[their \oiii\lin4363 direct method result]{Annibali:2019}) and a compilation of additional irregulars (\citealt[after removal of NGC~4449, for which agreement with the slope measured by \citealt{Annibali:2017} is found]{Pilyugin:2015}). While this comparison is potentially affected by systematics due to the adoption of different abundance diagnostics, the following conclusions will remain largely unmodified. It should also be  
noted that \cite{Hernandez-Martinez:2009} interpreted the \hii\ region data for NGC~6822 as consistent with a spatially homogeneous metallicity distribution. Moreover, deeper observations of enlarged samples of \hii\ regions in selected galaxies
have recently measured gradients in galaxies previously reported as chemically homogeneous (see \citealt{Annibali:2015} \vs\ \citealt{Lee:2004} for the case of NGC~1705).

% ..................................................................................................................
\begin{figure}
	\center
	\includegraphics[width=1\columnwidth]{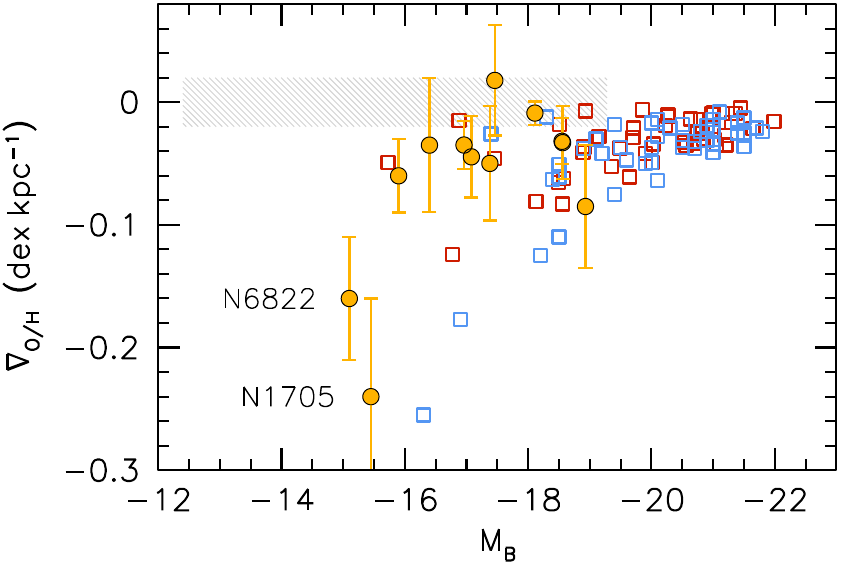}
	\caption{Abundance gradient slope \vs\ absolute $B$ magnitude for spiral galaxies from the LS sample (red squares), \citet[blue squares]{Ho:2015}, and dwarf irregular galaxies (orange points with error bars) from the literature
		(\citealt{Lee:2006b, Pilyugin:2015, Annibali:2015, Annibali:2017, Annibali:2019}). The shaded band represents the range in $M_B$ for irregular galaxies studied by \citet{Croxall:2009} and \citet{Haurberg:2013}, who found 
		their data consistent with chemical homogeneity.		
		\label{irregulars}}
\end{figure}
% ..................................................................................................................

Both oxygen and nitrogen appear to be homogeneously distributed, over $\sim$1~kpc spatial scales, in the interstellar medium of most dwarf irregular (\citealt{Croxall:2009, Haurberg:2013}) and starbursting blue compact dwarf galaxies (\citealt{Lagos:2014} and references therein). 
In the literature this is typically explained by a process of gas transport and mixing of the heavy elements, via supernova-driven metal-enriched winds, after cooling of the hot ejecta has occured on timescales $> 10^8$~yr (\citealt{Tenorio-Tagle:1996, Kobulnicky:1997, Legrand:2000}). 
An alternative mechanism, turbulent mixing, appears to work on comparable  timescales (\citealt{de-Avillez:2002, Yang:2012}).

The shaded band in Fig.~\ref{irregulars}, drawn with an arbitrary thickness, represents the $M_B$ range of the irregular galaxies studied by \citet{Croxall:2009} and \citet{Haurberg:2013} where the oxygen abundance analysis of at least three \hii\ regions is consistent with chemical homogeneity. However, the studies mentioned above indicate that chemical abundance gradients can develop in dwarf irregular galaxies, in the absence of the large-scale gravitational instabilities represented by  spiral arms, contrary to the suggestion made by \citet{Edmunds:1993}.
Moreover,  abundance gradients are measured for galaxies fainter than $M_B \simeq -17$, the limit proposed by \citet{Edmunds:1993}, down to at least $M_B \simeq -15$.

The data in Fig.~\ref{irregulars} suggest that (a) there is a considerable spread of gradients at low luminosity, (b) irregular galaxies can have gradients comparable to spirals, and (c) some irregulars are characterized by steep gradients, seemingly following the trend with luminosity and size established by small spiral galaxies.
This last point would suggest that irregular galaxies, at least down to a certain luminosity, may share the same inside-out growth mechanism that is supported by the steepening of the gradients of spirals with decreasing mass. 
However, this  contradicts the finding that the star-forming disk of low-mass irregular galaxies actually shrinks from the outside-in (\citealt{Zhang:2012}). This conclusion is based on the reddening of broad-band colors with increasing radius, \ie\ the shorter scale lengths at blue wavelengths compared to those at red wavelengths, at least in the outer disks.

The detection of significant metallicity gradients in irregular galaxies has been linked by \citet{Pilyugin:2015} to the presence of {\em steep} inner surface brightness profiles, whereas galaxies with {\em flat} inner light profiles exhibit flat metallicity gradients. 
Examination of Fig.~2 and 3 of \citet{Pilyugin:2015} suggests that the differentiation could be made instead between
'upbending' (Type ~III -- \citealt{Pohlen:2006}) and `downbending' (Type~II) or FI (flat or 'inverted', seen only in irregulars -- \citealt{Herrmann:2013}) surface brightness profiles\footnote{Only a minority of galaxies are characterized by pure exponential, Type~I profiles with no breaks (\eg\ \citealt{Pohlen:2006}).}. 
Galaxies with Type~III profiles tend to have significant abundance gradients, while those, more common, with Type~II profiles do not.
Objects with significant abundance gradients and not included in the study by \citet{Pilyugin:2015}, such as NGC~1705, NGC~4449 and DDO\,68, are indeed characterized by Type~III profiles (as inferred from the profiles published by \citealt{Munoz-Mateos:2009a} and 
\citealt{Pilyugin:2015} for the first two, and as reported by \citealt{Herrmann:2013} for the third). NGC~6822, on the other hand, represents an exception in having a Type~II profile (\citealt{Herrmann:2013}).

\citet{Herrmann:2013} have illustrated the varied and complex nature of multiband surface brightness profiles of dwarf irregular galaxies, highlighting similarities and differences with spiral galaxies.
They argued that differences in surface brightness profiles among dwarf galaxies reflect variations in star formation histories, with Type~III profiles resulting from enhanced central star formation and Type~II profiles from central star formation depletion, over timescales likely on the order of several hundred Myr. 
This picture is essentially confirmed by adding color information, with Type~III galaxies having inner bluer colors than Type~IIs and inner light profiles becoming redder with increasing radii (\citealt{Herrmann:2016}).

I speculate that the presence of significant radial metallicity gradients in Type~III profile irregular galaxies is consistent
with the notion of an inner star formation that is enhanced, on timescales that can be on the order of $\sim$100~Myr, when compared with the average star formation rate measured over the course of the lifetime of a galaxy.
There must be other factors playing a role, however. Among blue compact dwarf galaxies, which 
have experienced a star formation enhancement in recent times (\citealt{Zhang:2012}), only NGC~1705 is currently
known to have a measurable abundance gradient (\citealt{Annibali:2015}) (the other 
blue compact dwarf galaxies with homogeneous metallicity determinations are also much further away -- see \citealt{Lagos:2013} -- so perhaps spatial resolution is key in resolving their metallicity gradients).

Data on  star formation histories supports the idea that some of the irregular systems included in Fig.~\ref{irregulars}
have recently experienced an enhanced inner star formation activity.
For example, the star formation rate of NGC~6822 (although characterized by a Type~II light profile) has increased in its central region in the recent past according to resolved stellar studies (\citealt{Gallart:1996, Wyder:2001, Cannon:2012}).
Similarly, another non-bursting system, DDO~68, experienced a phase of enhanced inner star formation during the last few hundred  Myr, peaking between 10 and 100~Myr ago (\citealt{Sacchi:2016}).
The balance between the characteristic timescales for `recent' inner star formation enhancement (\eg\ time between separate bursts) and metal mixing could explain why some irregular galaxies have an abundance gradient and others (the majority?) do not. If the star formation `differential' between inner and outer regions subsides for a long period of time, these gradients should eventually disappear.

%==============================================================================================================
\section{Summary}
This work has presented long-slit spectroscopic data of four small, low-mass late-type spiral galaxies. Such galaxies are severely underrepresented in extragalactic abundance gradient studies. The spatial distribution of metals has been investigated for only a handful of spiral galaxies with disk scale lengths $< 1.5-2$~kpc and/or stellar masses below $10^9$~\msun.
Yet investigations in this regime, adjoining the regime of dwarf irregular galaxies, are important to test variations of abundance gradients as a function of stellar mass, and predictions made by models and numerical simulations of galaxy evolution. Further verification of some of the trends highlighted here, for example how the slope of the abundance gradients depend on scale length, luminosity and surface brightness profile type, will certainly benefit from additional observations of nearby small galaxies.

The oxygen abundances measured for the \hii\ regions in the outer disk of NGC~1058 confirm the flattening of the radial metallicity gradient taking place around the isophotal radius found for similar, extended disk galaxies (\citealt{Bresolin:2017}).

I have combined the new observations presented here with a sample of spiral galaxies drawn from the compilation by \citet{Pilyugin:2014}, in which the chemical abundances were derived from slit spectroscopy (rather than IFUs). The properties of the gas-phase metallicity gradients of this Long Slit sample are largely in agreement, at least qualitatively, with those of larger IFU studies, such as those by \citet{Sanchez-Menguiano:2016} and \citet{Sanchez-Menguiano:2018}. These data are then compared to predictions from recent cosmological simulations of galaxy evolution, finding that often the simulations fail to reproduce the mean steepening of the gradients (in dex\,kpc$^{-1}$) with decreasing stellar mass, or do not extend to sufficiently small stellar masses for a meaningful comparison. I advocate the use of nearby small galaxies for tests of galaxy simulations, 
for which extensions to stellar masses down to \logm\,$\sim$\,8.0 would be necessary.

The comparison with the outside-in chemical evolution models of \citet{Prantzos:2000} show qualitative agreement, but the models predict systematically steeper abundance slopes than observed using a variety of abundance diagnostics. It is argued that this may be due to
the importance of processes that are neglected in the models, such as the outflow of gas or radial mixing. The significant scatter observed for the abundance gradient slope at a given scale length, mass or luminosity can be related to variations in star formation efficiency or stellar feedback. 

Lastly, assembling abundance gradient data for spiral galaxies extending to low intrinsic luminosities and similar data for 
a handful of irregular galaxies suggests that dwarf galaxies characterized by a recent star formation enhancement in the inner disk could be those where we detect also chemical abundance gradients.
Contrary to spiral galaxies, these gradients would have a transitory nature.

% ..................................................................................................................

\bigskip
\bigskip
\noindent

\section*{Acknowledgments}
I thank the anonymous referee for suggestions that improved the
clarity of this work.	
Based on observations collected at the Gemini Observatory, which is operated by the 
Association of Universities for Research in Astronomy, Inc., under a cooperative agreement 
with the NSF on behalf of the Gemini partnership: the National Science Foundation (United 
States), the Science and Technology Facilities Council (United Kingdom), the 
National Research Council (Canada), CONICYT (Chile), the Australian Research Council (Australia), 
Minist\'{e}rio da Ci\^{e}ncia e Tecnologia (Brazil) 
and Ministerio de Ciencia, Tecnolog\'{i}a e Innovaci\'{o}n Productiva (Argentina). 
The program reference numbers are: GN-2011-B-Q-84 and GN-2017A-Q-25.
I acknowledge the usage of the HyperLeda database (http://leda.univ-lyon1.fr).
This research has made use of the NASA/IPAC Extragalactic Database (NED), which is operated by the Jet Propulsion Laboratory, California Institute of Technology, under contract with the National Aeronautics and Space Administration.
This research has made use of the SIMBAD database, operated at CDS, Strasbourg, France. 
%\clearpage
\bibliographystyle{mnras}
%\bibliography{Papers}
\bibliography{References}

\appendix
\section{Long slit sample properties}\label{Appendix:A}
This table summarizes the main properties of the long slit sample used in this paper. 
All distance-dependent quantities have been scaled by the distance adopted here.

The columns are defined as follows:
(1) galaxy identification; (2) morphological classification from the RC3 catalog (\citealt{de-Vaucouleurs:1991}), as presented in NED; (3) revisited distance in Mpc; (4) absolute $B$ magnitude derived from the extinction-corrected $B$ magnitude from HyperLeda;
(5) stellar mass; (6) disk scale length in kpc extracted from the literature; 
(7) \rtf\ in kpc from \citet{Pilyugin:2015}; (8) slope of the O/H gradient 
in dex\,kpc$^{-1}$ from \citet{Pilyugin:2015}; (9) 
slope of the O/H gradient in dex\,\re$^{-1}$.

The abundance gradient slopes of the last four objects were derived from the Gemini/GMOS data, and their other parameters are taken from 
Table~\ref{tab:sample}.

% ..................................................................................................................
%  TABLE: galaxies
\begin{table*}
	\centering
	\caption{Long slit galaxy sample.}\label{tab:galaxies}
	\begin{tabular}{lcccccccc}
		\hline
ID	&	      Type    &    D (Mpc)          & M$_B$  & log(M/M$_\odot$) &  $r_d$ (kpc) &  \rtf\ (kpc)   &    \slope\ (dex\,kpc$^{-1}$)  &   \slope\ (dex\,$r_e^{-1}$) \\
(1) &         (2)     &   (3)              & (4)    & (5)            & (6)     & (7)     & (8)    & (9)      \\	
\hline
NGC 300    &   Sd     &         2.02  & $-$18.13 &  9.43 &    2.19  &   6.43  & $-$0.081 $\pm$  0.006  & $-$0.297 $\pm$  0.024  \\ 
NGC 450    &   SABcd  &         17.5  & $-$18.91 &  9.57 &    2.92  &   7.84  & $-$0.036 $\pm$  0.007  & $-$0.178 $\pm$  0.033  \\ 
NGC 598    &   Scd    &         0.86  & $-$18.89 &  9.59 &    1.94  &   8.81  & $-$0.041 $\pm$  0.005  & $-$0.133 $\pm$  0.017  \\ 
NGC 753    &   SABbc  &         42.1  & $-$20.80 &  10.42 &    2.49  &  15.37  & $-$0.014 $\pm$  0.007  & $-$0.058 $\pm$  0.018  \\ 
NGC 925    &   SABd   &          9.1  & $-$20.03 &  10.00 &    4.59  &  13.89  & $-$0.034 $\pm$  0.003  & $-$0.262 $\pm$  0.020  \\ 
NGC 1068   &   Sb     &         12.3  & $-$20.98 &  10.82 &    7.65  &  12.67  & $-$0.010 $\pm$  0.004  & $-$0.122 $\pm$  0.063  \\ 
NGC 1097   &   SBb    &         15.8  & $-$21.30 &  10.93 &    6.75  &  21.42  & $-$0.017 $\pm$  0.003  & $-$0.188 $\pm$  0.027  \\ 
NGC 1232   &   SABc   &         15.0  & $-$20.56 &  10.46 &    5.26  &  16.17  & $-$0.035 $\pm$  0.004  & $-$0.313 $\pm$  0.027  \\ 
NGC 1313   &   SBd    &         4.31  & $-$19.15 &  9.50 &    1.34  &   5.71  & $-$0.028 $\pm$  0.009  & $-$0.063 $\pm$  0.020  \\ 
NGC 1365   &   SBb    &         17.2  & $-$21.36 &  10.83 &    7.09  &  28.11  & $-$0.009 $\pm$  0.001  & $-$0.107 $\pm$  0.012  \\ 
NGC 1672   &   SBb    &         11.9  & $-$20.23 &  10.29 &    2.48  &  11.45  & $-$0.019 $\pm$  0.006  & $-$0.079 $\pm$  0.018  \\ 
NGC 2336   &   SABbc  &         33.4  & $-$21.99 &  10.94 &   11.41  &  24.38  & $-$0.016 $\pm$  0.002  & $-$0.306 $\pm$  0.042  \\ 
NGC 2403   &   SABcd  &         3.13  & $-$19.35 &  9.61 &    1.48  &   9.97  & $-$0.053 $\pm$  0.004  & $-$0.131 $\pm$  0.011  \\ 
NGC 2835   &   SBc    &          8.7  & $-$19.49 &  9.60 &    2.59  &   8.41  & $-$0.037 $\pm$  0.007  & $-$0.163 $\pm$  0.027  \\ 
NGC 2903   &   SABbc  &          9.3  & $-$21.02 &  10.63 &    2.90  &  17.09  & $-$0.031 $\pm$  0.004  & $-$0.149 $\pm$  0.021  \\ 
NGC 2997   &   SABc   &         11.3  & $-$21.06 &  10.55 &    4.59  &  14.62  & $-$0.034 $\pm$  0.008  & $-$0.265 $\pm$  0.059  \\ 
NGC 3198   &   SBc    &         13.7  & $-$20.74 &  10.25 &    3.86  &  16.93  & $-$0.025 $\pm$  0.004  & $-$0.163 $\pm$  0.027  \\ 
NGC 3344   &   SABbc  &          9.8  & $-$19.64 &  10.04 &    2.51  &  10.11  & $-$0.061 $\pm$  0.004  & $-$0.257 $\pm$  0.027  \\ 
NGC 3351   &   SBb    &          9.3  & $-$19.71 &  10.39 &    2.64  &  10.06  & $-$0.021 $\pm$  0.003  & $-$0.093 $\pm$  0.012  \\ 
NGC 3359   &   SBc    &         18.9  & $-$20.63 &  10.19 &    4.85  &  19.89  & $-$0.014 $\pm$  0.005  & $-$0.111 $\pm$  0.058  \\ 
NGC 3621   &   Sd     &          6.5  & $-$20.02 &  10.04 &    2.00  &   9.30  & $-$0.049 $\pm$  0.004  & $-$0.166 $\pm$  0.011  \\ 
NGC 4254   &   Sc     &         13.9  & $-$20.54 &  10.33 &    2.63  &  10.83  & $-$0.033 $\pm$  0.003  & $-$0.144 $\pm$  0.014  \\ 
NGC 4303   &   SABbc  &         17.6  & $-$21.21 &  10.57 &    3.51  &  16.54  & $-$0.035 $\pm$  0.004  & $-$0.206 $\pm$  0.034  \\ 
NGC 4321   &   SABbc  &         14.3  & $-$20.95 &  10.76 &    4.11  &  15.45  & $-$0.014 $\pm$  0.003  & $-$0.098 $\pm$  0.018  \\ 
NGC 4395   &   Sm     &         4.76  & $-$18.51 &  9.21 &    3.43  &   9.13  & $-$0.018 $\pm$  0.009  & $-$0.104 $\pm$  0.058  \\ 
NGC 5033   &   Sc     &         19.1  & $-$21.32 &  10.57 &    6.69  &  29.69  & $-$0.009 $\pm$  0.003  & $-$0.105 $\pm$  0.036  \\ 
NGC 5068   &   SABcd  &          5.4  & $-$18.57 &  9.62 &    2.02  &   5.74  & $-$0.062 $\pm$  0.017  & $-$0.210 $\pm$  0.059  \\ 
NGC 5236   &   SABc   &         4.61  & $-$20.82 &  10.58 &    2.75  &   8.65  & $-$0.026 $\pm$  0.003  & $-$0.118 $\pm$  0.015  \\ 
NGC 5248   &   SABbc  &         13.6  & $-$20.27 &  10.34 &    2.61  &  12.16  & $-$0.012 $\pm$  0.006  & $-$0.051 $\pm$  0.016  \\ 
NGC 5457   &   SABcd  &          6.7  & $-$20.84 &  10.32 &    5.86  &  28.10  & $-$0.030 $\pm$  0.001  & $-$0.295 $\pm$  0.010  \\ 
NGC 5668   &   Sd     &         23.6  & $-$19.90 &  10.04 &    2.77  &  11.35  & $-$0.042 $\pm$  0.005  & $-$0.194 $\pm$  0.022  \\ 
NGC 6384   &   SABbc  &         27.5  & $-$21.65 &  10.79 &    6.14  &  24.70  & $-$0.021 $\pm$  0.003  & $-$0.221 $\pm$  0.030  \\ 
NGC 6744   &   SABbc  &          9.2  & $-$21.19 &  10.68 &    5.46  &  26.59  & $-$0.027 $\pm$  0.002  & $-$0.250 $\pm$  0.020  \\ 
NGC 6946   &   SABcd  &          6.1  & $-$20.69 &  10.20 &    3.82  &  10.16  & $-$0.033 $\pm$  0.010  & $-$0.213 $\pm$  0.063  \\ 
NGC 7518   &   SABa   &         59.4  & $-$20.27 &  10.49 &    3.89  &  12.22  & $-$0.010 $\pm$  0.004  & $-$0.068 $\pm$  0.028  \\ 
NGC 7529   &   Sbc    &         64.9  & $-$19.70 &  9.85 &    2.55  &   8.03  & $-$0.029 $\pm$  0.008  & $-$0.124 $\pm$  0.037  \\ 
NGC 7591   &   SBbc   &         59.4  & $-$21.01 &  10.75 &    5.94  &  16.85  & $-$0.009 $\pm$  0.002  & $-$0.086 $\pm$  0.017  \\ 
NGC 7678   &   SABc   &         50.1  & $-$21.45 &  10.64 &    5.25  &  17.09  & $-$0.005 $\pm$  0.005  & $-$0.041 $\pm$  0.041  \\ 
NGC 7793   &   Sd     &         3.44  & $-$18.49 &  9.38 &    1.10  &   4.67  & $-$0.066 $\pm$  0.010  & $-$0.121 $\pm$  0.019  \\ 
IC 5309    &   Sb     &         47.9  & $-$19.86 &  10.31 &    2.92  &   9.39  & $-$0.006 $\pm$  0.007  & $-$0.030 $\pm$  0.033  \\ 
UGC 2345   &   SBm    &         14.3  & $-$16.77 &  8.80 &    1.93  &   7.19  & $-$0.124 $\pm$  0.012  & $-$0.402 $\pm$  0.027  \\ 
UGC 10445  &   Scd    &         28.7  & $-$18.93 &  9.36 &    2.83  &  11.50  & $-$0.007 $\pm$  0.007  & $-$0.034 $\pm$  0.046  \\ 
\hline
NGC 1058   &   Sc     &          9.1  & $-$18.56 &  9.53 &    1.18  &   4.01  & $-$0.083 $\pm$  0.008  & $-$0.165 $\pm$  0.016  \\ 
UGC 7490   &   Sm     &          9.0  & $-$16.88 &  8.52 &    1.60  &   3.28  & $-$0.015 $\pm$  0.018  & $-$0.040 $\pm$  0.048  \\ 
NGC 4523   &   SABm   &         16.8  & $-$17.45 &  8.64 &    2.52  &   6.49  & $-$0.046 $\pm$  0.011  & $-$0.195 $\pm$  0.047  \\ 
NGC 4707   &   Sm     &          6.5  & $-$15.73 &  8.33 &    1.26  &   1.95  & $-$0.049 $\pm$  0.062  & $-$0.103 $\pm$  0.131  \\ 
		\hline
	\end{tabular}\\
\end{table*}
% ..................................................................................................................

\section{Reddening-corrected line fluxes and oxygen abundances tables}\label{Appendix:B}
The tables present reddening-corrected line fluxes, normalized to I(\hbeta)\,=\,100, and oxygen abundances 12\,+\,log(O/H) for the four galaxies observed with Gemini/GMOS.
 
% ..................................................................................................................
%  TABLES: FLUXES
\begin{table*}
	\centering
	\begin{minipage}{16.cm}
		\centering
		\caption{NGC~1058}\label{tab:fluxes1058}
		\begin{tabular}{lcccccccc}
			\hline
ID	&	R.A.		&	Dec.		&	$r/r_{25}$		&	\oii\		&	\oiii\ 		& \nii\		&	\sii\			&	  12+log(O/H)\\
&	(J2000.0)	&	(J2000.0)	&				    &	3727  	    &	5007		&	6583		&	6731+6717	    &	  	\\
(1) &   (2)         & (3)           & (4)               & (5)           & (6)           & (7)           & (8)               & (9)   \\	
\hline
1 &   02 43 40.4  & $37$ 23 26.70  & 2.55 &    313 $\pm$   76 &     141 $\pm$   13 &      25 $\pm$    4 &      63 $\pm$    6 &   8.22 $\pm$ 0.09 \\
2 &   02 43 33.0  & $37$ 23 08.88  & 1.96 &    323 $\pm$   86 &      78 $\pm$   10 &      34 $\pm$    5 &      55 $\pm$    6 &   8.30 $\pm$ 0.09 \\
3 &   02 43 35.2  & $37$ 22 47.68  & 1.81 &    211 $\pm$   27 &     257 $\pm$   13 &      22 $\pm$    2 &      33 $\pm$    2 &   8.22 $\pm$ 0.04 \\
4 &   02 43 42.0  & $37$ 22 30.58  & 2.16 &    224 $\pm$   56 &     380 $\pm$   59 &      19 $\pm$    4 &      41 $\pm$    6 &   8.16 $\pm$ 0.04 \\
5 &   02 43 40.9  & $37$ 22 19.92  & 1.97 &    422 $\pm$   96 &     172 $\pm$   18 &      28 $\pm$    4 &      42 $\pm$    5 &   8.23 $\pm$ 0.11 \\
6 &   02 43 39.5  & $37$ 22 08.64  & 1.74 &    \nodata        &      97 $\pm$   14 &      37 $\pm$    6 &      51 $\pm$    7 &   8.36 $\pm$ 0.05 \\
7 &   02 43 42.4  & $37$ 21 59.14  & 1.97 &    177 $\pm$   20 &     332 $\pm$   16 &      17 $\pm$    1 &      31 $\pm$    2 &   8.13 $\pm$ 0.04 \\
8 &   02 43 40.6  & $37$ 21 52.10  & 1.72 &    \nodata        &     172 $\pm$   16 &      34 $\pm$    4 &      55 $\pm$    6 &   8.31 $\pm$ 0.03 \\
9 &   02 43 42.2  & $37$ 21 45.93  & 1.86 &    272 $\pm$   28 &     201 $\pm$   10 &      28 $\pm$    2 &      47 $\pm$    3 &   8.25 $\pm$ 0.05 \\
10 &   02 43 34.1  & $37$ 21 26.28  & 0.88 &    197 $\pm$   16 &     176 $\pm$    8 &      40 $\pm$    3 &      41 $\pm$    2 &   8.35 $\pm$ 0.03 \\
11 &   02 43 40.2  & $37$ 21 21.68  & 1.49 &    \nodata        &     575 $\pm$   40 &      25 $\pm$    3 &      50 $\pm$    4 &   8.21 $\pm$ 0.05 \\
12 &   02 43 33.0  & $37$ 21 07.56  & 0.61 &    195 $\pm$   13 &     124 $\pm$    6 &      63 $\pm$    4 &      41 $\pm$    2 &   8.46 $\pm$ 0.04 \\
13 &   02 43 31.4  & $37$ 21 03.92  & 0.46 &    158 $\pm$   11 &     112 $\pm$    5 &      58 $\pm$    4 &      35 $\pm$    2 &   8.46 $\pm$ 0.04 \\
14 &   02 43 34.1  & $37$ 21 03.90  & 0.69 &    264 $\pm$   28 &     140 $\pm$    7 &      72 $\pm$    5 &     107 $\pm$    5 &   8.46 $\pm$ 0.06 \\
15 &   02 43 31.6  & $37$ 20 55.75  & 0.39 &    122 $\pm$   11 &      22 $\pm$    1 &      94 $\pm$    6 &      76 $\pm$    4 &   8.61 $\pm$ 0.08 \\
16 &   02 43 31.7  & $37$ 20 47.25  & 0.31 &    106 $\pm$   12 &      28 $\pm$    2 &      94 $\pm$    6 &      69 $\pm$    4 &   8.61 $\pm$ 0.07 \\
17 &   02 43 30.4  & $37$ 20 39.63  & 0.14 &     88 $\pm$   11 &      19 $\pm$    2 &      92 $\pm$    6 &      44 $\pm$    2 &   8.63 $\pm$ 0.08 \\
18 &   02 43 29.5  & $37$ 20 36.50  & 0.11 &     89 $\pm$   17 &      18 $\pm$    1 &      90 $\pm$    6 &      42 $\pm$    2 &   8.63 $\pm$ 0.08 \\
19 &   02 43 33.4  & $37$ 20 35.55  & 0.46 &    179 $\pm$   12 &     114 $\pm$    5 &      77 $\pm$    5 &      41 $\pm$    2 &   8.51 $\pm$ 0.05 \\
20 &   02 43 31.3  & $37$ 20 35.52  & 0.19 &    \nodata        &      15 $\pm$    2 &     104 $\pm$    7 &      73 $\pm$    4 &   8.65 $\pm$ 0.11 \\
21 &   02 43 35.9  & $37$ 20 25.66  & 0.78 &    206 $\pm$   27 &      87 $\pm$    6 &      66 $\pm$    5 &      54 $\pm$    3 &   8.48 $\pm$ 0.04 \\
22 &   02 43 31.9  & $37$ 20 17.05  & 0.29 &    \nodata        &      26 $\pm$    3 &      91 $\pm$    7 &      35 $\pm$    2 &   8.61 $\pm$ 0.08 \\
23 &   02 43 28.4  & $37$ 20 17.01  & 0.26 &    116 $\pm$    8 &      24 $\pm$    1 &     103 $\pm$    6 &      80 $\pm$    4 &   8.63 $\pm$ 0.08 \\
24 &   02 43 20.5  & $37$ 20 15.83  & 1.26 &    207 $\pm$   24 &     306 $\pm$   14 &      26 $\pm$    2 &      30 $\pm$    2 &   8.26 $\pm$ 0.04 \\
25 &   02 43 30.9  & $37$ 20 13.68  & 0.22 &    113 $\pm$   13 &      16 $\pm$    1 &      88 $\pm$    6 &      47 $\pm$    2 &   8.62 $\pm$ 0.09 \\
26 &   02 43 36.1  & $37$ 20 07.77  & 0.84 &    132 $\pm$   15 &     262 $\pm$   12 &      28 $\pm$    2 &      23 $\pm$    1 &   8.31 $\pm$ 0.05 \\
27 &   02 43 28.7  & $37$ 19 53.29  & 0.47 &    158 $\pm$    9 &      54 $\pm$    2 &      87 $\pm$    6 &      56 $\pm$    3 &   8.56 $\pm$ 0.05 \\
28 &   02 43 29.4  & $37$ 19 46.52  & 0.52 &    118 $\pm$   17 &      29 $\pm$    2 &      83 $\pm$    6 &      63 $\pm$    3 &   8.58 $\pm$ 0.07 \\
29 &   02 43 29.3  & $37$ 19 36.38  & 0.64 &    198 $\pm$   13 &      57 $\pm$    3 &      82 $\pm$    5 &      65 $\pm$    3 &   8.53 $\pm$ 0.05 \\
30 &   02 43 33.7  & $37$ 19 36.04  & 0.80 &    179 $\pm$   21 &     313 $\pm$   15 &      38 $\pm$    3 &      41 $\pm$    2 &   8.33 $\pm$ 0.05 \\
31 &   02 43 36.5  & $37$ 19 18.41  & 1.20 &    188 $\pm$   65 &     435 $\pm$   29 &      19 $\pm$    2 &      26 $\pm$    2 &   8.18 $\pm$ 0.03 \\
32 &   02 43 29.4  & $37$ 19 12.40  & 0.92 &    166 $\pm$   25 &     270 $\pm$   14 &      31 $\pm$    2 &      35 $\pm$    2 &   8.30 $\pm$ 0.04 \\
33 &   02 43 21.6  & $37$ 19 09.62  & 1.47 &    355 $\pm$   61 &     134 $\pm$    8 &      29 $\pm$    3 &      47 $\pm$    4 &   8.25 $\pm$ 0.09 \\
34 &   02 43 29.3  & $37$ 18 57.34  & 1.10 &    203 $\pm$   12 &     271 $\pm$   12 &      36 $\pm$    2 &      51 $\pm$    2 &   8.31 $\pm$ 0.04 \\
35 &   02 43 43.3  & $37$ 18 41.80  & 2.16 &    197 $\pm$   19 &     514 $\pm$   25 &      13 $\pm$    1 &      29 $\pm$    2 &   8.11 $\pm$ 0.05 \\
36 &   02 43 28.7  & $37$ 18 40.29  & 1.32 &    232 $\pm$   35 &     268 $\pm$   14 &      30 $\pm$    2 &      39 $\pm$    3 &   8.27 $\pm$ 0.04 \\
37 &   02 43 23.2  & $37$ 18 14.94  & 1.85 &    \nodata        &     127 $\pm$   10 &      29 $\pm$    3 &      58 $\pm$    5 &   8.30 $\pm$ 0.05 \\
38 &   02 43 23.8  & $37$ 17 57.51  & 2.00 &    194 $\pm$   44 &     367 $\pm$   20 &      15 $\pm$    2 &      29 $\pm$    2 &   8.12 $\pm$ 0.04 \\
39 &   02 43 29.2  & $37$ 17 40.99  & 2.02 &    286 $\pm$   41 &      44 $\pm$    4 &      28 $\pm$    2 &      77 $\pm$    5 &   8.28 $\pm$ 0.13 \\
40 &   02 43 27.0  & $37$ 17 26.98  & 2.22 &    453 $\pm$   81 &     156 $\pm$    9 &      20 $\pm$    2 &      40 $\pm$    3 &   8.15 $\pm$ 0.14 \\
			\hline
		\end{tabular}
	\end{minipage}	
\end{table*}

\begin{table*}
	\centering
	\begin{minipage}{16.cm}
		\centering
		\caption{UGC~7490}\label{tab:fluxes7490}
		\begin{tabular}{lcccccccc}
			\hline
ID	&	R.A.		&	Dec.		&	$r/r_{25}$		&	\oiii\ 		& \nii\		&	\sii\			&	12+log(O/H) \\
&	(J2000.0)	&	(J2000.0)	&				    &	5007		&	6583		&	6731+6717	    &				\\
(1) &   (2)         & (3)           & (4)               & (5)           & (6)           & (7)           & (8)            \\	
\hline
1 &   12 24 37.4  & $70$ 19 35.87  & 0.88 &    260 $\pm$   12 &      20 $\pm$    1 &      42 $\pm$    2 &   8.21 $\pm$ 0.01 \\
2 &   12 24 30.9  & $70$ 19 44.66  & 0.44 &    205 $\pm$   10 &      31 $\pm$    2 &      76 $\pm$    3 &   8.28 $\pm$ 0.04 \\
4 &   12 24 27.0  & $70$ 19 26.39  & 0.47 &    367 $\pm$   18 &      21 $\pm$    1 &      41 $\pm$    2 &   8.20 $\pm$ 0.02 \\
5 &   12 24 25.1  & $70$ 18 46.40  & 0.99 &    144 $\pm$    7 &      24 $\pm$    1 &      51 $\pm$    2 &   8.27 $\pm$ 0.04 \\
6 &   12 24 20.2  & $70$ 20 31.73  & 0.53 &    250 $\pm$   14 &      40 $\pm$    3 &     104 $\pm$    5 &   8.30 $\pm$ 0.05 \\
7 &   12 24 22.1  & $70$ 19 58.77  & 0.21 &    196 $\pm$   13 &      53 $\pm$    4 &     144 $\pm$    7 &   8.36 $\pm$ 0.07 \\
8 &   12 24 26.9  & $70$ 21 00.66  & 0.80 &    370 $\pm$   18 &      16 $\pm$    1 &      41 $\pm$    2 &   8.15 $\pm$ 0.03 \\
9 &   12 24 30.9  & $70$ 19 18.95  & 0.67 &    171 $\pm$    8 &      31 $\pm$    2 &      74 $\pm$    3 &   8.28 $\pm$ 0.04 \\
10 &   12 24 20.4  & $70$ 19 00.02  & 0.88 &    147 $\pm$   11 &      32 $\pm$    2 &      62 $\pm$    4 &   8.31 $\pm$ 0.04 \\
12 &   12 24 21.8  & $70$ 20 50.68  & 0.70 &    349 $\pm$   19 &      20 $\pm$    1 &      40 $\pm$    2 &   8.19 $\pm$ 0.02 \\
			\hline
		\end{tabular}
	\end{minipage}	
\end{table*}

\begin{table*}
	\centering
	\begin{minipage}{16.cm}
		\centering
		\caption{NGC~4523}\label{tab:fluxes4523}
		\begin{tabular}{lcccccccc}
			\hline
ID	&	R.A.		&	Dec.		&	$r/r_{25}$		&	\oiii\ 		& \nii\		&	\sii\			&	12+log(O/H) \\
&	(J2000.0)	&	(J2000.0)	&				    &	5007		&	6583		&	6731+6717	    &				\\
(1) &   (2)         & (3)           & (4)               & (5)           & (6)           & (7)           & (8)            \\	
\hline
1 &   12 33 47.4  & $15$ 10 49.33  & 0.60 &    278 $\pm$   13 &      17 $\pm$    1 &      32 $\pm$    1 &   8.15 $\pm$ 0.07 \\
2 &   12 33 47.3  & $15$ 10 40.30  & 0.50 &    354 $\pm$   16 &      18 $\pm$    1 &      39 $\pm$    2 &   8.16 $\pm$ 0.03 \\
3 &   12 33 47.3  & $15$ 10 32.34  & 0.40 &    248 $\pm$   12 &      24 $\pm$    1 &      37 $\pm$    2 &   8.25 $\pm$ 0.02 \\
4 &   12 33 48.3  & $15$ 10 15.17  & 0.13 &     52 $\pm$    3 &      46 $\pm$    3 &      84 $\pm$    3 &   8.41 $\pm$ 0.09 \\
5 &   12 33 47.8  & $15$ 10 01.67  & 0.07 &    288 $\pm$   13 &      40 $\pm$    2 &      68 $\pm$    3 &   8.31 $\pm$ 0.04 \\
6 &   12 33 46.9  & $15$ 09 48.96  & 0.30 &    153 $\pm$    7 &      26 $\pm$    1 &      32 $\pm$    1 &   8.29 $\pm$ 0.03 \\
7 &   12 33 48.8  & $15$ 09 52.78  & 0.27 &    335 $\pm$   17 &      23 $\pm$    1 &      36 $\pm$    2 &   8.22 $\pm$ 0.02 \\
8 &   12 33 47.7  & $15$ 09 31.75  & 0.44 &    261 $\pm$   16 &      27 $\pm$    2 &      47 $\pm$    2 &   8.25 $\pm$ 0.02 \\
9 &   12 33 43.6  & $15$ 09 20.41  & 1.04 &    386 $\pm$   17 &      15 $\pm$    1 &      48 $\pm$    2 &   8.14 $\pm$ 0.02 \\
10 &   12 33 52.7  & $15$ 10 54.41  & 1.11 &    214 $\pm$   16 &      17 $\pm$    2 &      37 $\pm$    2 &   8.14 $\pm$ 0.11 \\
11 &   12 33 46.5  & $15$ 09 44.35  & 0.40 &    280 $\pm$   13 &      31 $\pm$    2 &      60 $\pm$    2 &   8.27 $\pm$ 0.03 \\
12 &   12 33 52.0  & $15$ 10 07.72  & 0.87 &    114 $\pm$    7 &      21 $\pm$    2 &      47 $\pm$    2 &   8.25 $\pm$ 0.05 \\
13 &   12 33 45.3  & $15$ 09 37.40  & 0.64 &    253 $\pm$   14 &      33 $\pm$    2 &     101 $\pm$    5 &   8.27 $\pm$ 0.05 \\
			\hline
		\end{tabular}
	\end{minipage}
\end{table*}

\begin{table*}
	\centering
	\begin{minipage}{16.cm}
		\centering
		\caption{NGC~4707}\label{tab:fluxes4707}
		\begin{tabular}{lcccccccc}
			\hline
ID	&	R.A.		&	Dec.		&	$r/r_{25}$		&	\oiii\ 		& \nii\		&	\sii\			&	12+log(O/H) \\
&	(J2000.0)	&	(J2000.0)	&				    &	5007		&	6583		&	6731+6717	    &				\\
(1) &   (2)         & (3)           & (4)               & (5)           & (6)           & (7)           & (8)            \\	
\hline
1 &   12 48 25.8  & $51$ 10 28.64  & 0.74 &    378 $\pm$   17 &       9 $\pm$    0 &      25 $\pm$    1 &   8.04 $\pm$ 0.06 \\
2 &   12 48 26.7  & $51$ 10 16.42  & 0.76 &    202 $\pm$    9 &      13 $\pm$    1 &      38 $\pm$    2 &   8.09 $\pm$ 0.14 \\
3 &   12 48 26.8  & $51$ 10 18.78  & 0.78 &    259 $\pm$   12 &      11 $\pm$    1 &      42 $\pm$    2 &   8.08 $\pm$ 0.11 \\
4 &   12 48 23.3  & $51$ 10 06.48  & 0.24 &    176 $\pm$    9 &      19 $\pm$    1 &      50 $\pm$    2 &   8.16 $\pm$ 0.14 \\
5 &   12 48 24.7  & $51$ 09 42.62  & 0.46 &    630 $\pm$   28 &       8 $\pm$    0 &      27 $\pm$    1 &   8.03 $\pm$ 0.04 \\
6 &   12 48 20.5  & $51$ 09 26.20  & 0.57 &    293 $\pm$   16 &       9 $\pm$    1 &      30 $\pm$    2 &   8.05 $\pm$ 0.11 \\
7 &   12 48 20.4  & $51$ 09 30.10  & 0.54 &    444 $\pm$   21 &       7 $\pm$    0 &      24 $\pm$    1 &   8.01 $\pm$ 0.04 \\
8 &   12 48 25.5  & $51$ 09 07.80  & 1.08 &    148 $\pm$    9 &      14 $\pm$    1 &      44 $\pm$    2 &   8.09 $\pm$ 0.19 \\
9 &   12 48 24.1  & $51$ 10 00.31  & 0.25 &     72 $\pm$    6 &      22 $\pm$    2 &      58 $\pm$    3 &   8.27 $\pm$ 0.09 \\			
			\hline
		\end{tabular}
	\end{minipage}
\end{table*}

\bsp	% typesetting comment
\label{lastpage}
\end{document}